\pdfoutput=1
\documentclass[10pt,twocolumn,letterpaper]{article}

\usepackage[margin=0.9in]{geometry}
\usepackage{epsfig}
\usepackage{graphicx}
\usepackage{amsmath}
\usepackage{amssymb}
\usepackage{amsthm}
\usepackage{nicefrac}
\usepackage[]{algorithm}
\usepackage{algpseudocode}
\usepackage{siunitx}
\algnewcommand{\LineComment}[1]{\State \(\triangleright\) #1}

\theoremstyle{theorem}
\theoremstyle{definition}
\theoremstyle{remark}

\algrenewcommand\algorithmicrequire{\textbf{Input:}}
\algrenewcommand\algorithmicensure{\textbf{Output:}}
\newcounter{MYtempeqncnt}

\newcommand*{\nolink}[1]{%
  {\protect\NoHyper#1\protect\endNoHyper}%
}

\global\hyphenpenalty=100
\begin{document}

\newcommand{\mytitle}{\vspace{-0.1in}High Flux Passive Imaging with Single-Photon Sensors\vspace{-0.1in}}
\title{\mytitle}

\author{
  Atul Ingle \,\,\,\,
  Andreas Velten$^\dagger$ \,\,\,\,
  Mohit Gupta$^\dagger$ \\
  {\tt\small \{ingle,velten,mgupta37\}@wisc.edu}\vspace{2pt} \\
  University of Wisconsin-Madison
}

\maketitle
\renewcommand*{\thefootnote}{$\dagger$}
\setcounter{footnote}{1}
\footnotetext{Equal contribution.\newline
  This research was supported in part by ONR grants N00014-15-1-2652 and
N00014-16-1-2995 and DARPA grant HR0011-16-C-0025.}
\renewcommand*{\thefootnote}{\arabic{footnote}}
\setcounter{footnote}{0}
\thispagestyle{empty}

\begin{abstract}
Single-photon avalanche diodes (SPADs) are an emerging technology with a unique
capability of capturing individual photons with high timing precision. SPADs
are being used in several active imaging systems (e.g., fluorescence lifetime
microscopy and LiDAR), albeit mostly limited to low photon flux settings. We
propose \textnormal{passive free-running SPAD (PF-SPAD)} imaging, an imaging
modality that uses SPADs for capturing 2D intensity images with unprecedented
dynamic range under ambient lighting, without any active light source. Our key
observation is that the precise inter-photon timing measured by a SPAD can be
used for estimating scene brightness under ambient lighting conditions, even
for very bright scenes. We develop a theoretical model for PF-SPAD imaging, and
derive a scene brightness estimator based on the average time of darkness
between successive photons detected by a PF-SPAD pixel. Our key insight is that
due to the stochastic nature of photon arrivals, this estimator does not suffer
from a hard saturation limit.  Coupled with high sensitivity at low flux, this
enables a PF-SPAD pixel to measure a wide range of scene brightnesses, from
very low to very high, thereby achieving extreme dynamic range. We demonstrate
an improvement of over 2 orders of magnitude over conventional sensors by
imaging scenes spanning a dynamic range of $10^6:1$.

\end{abstract}

\section{Introduction}
\begin{figure*}[!ht]
  \centering \includegraphics[width=0.95\textwidth]{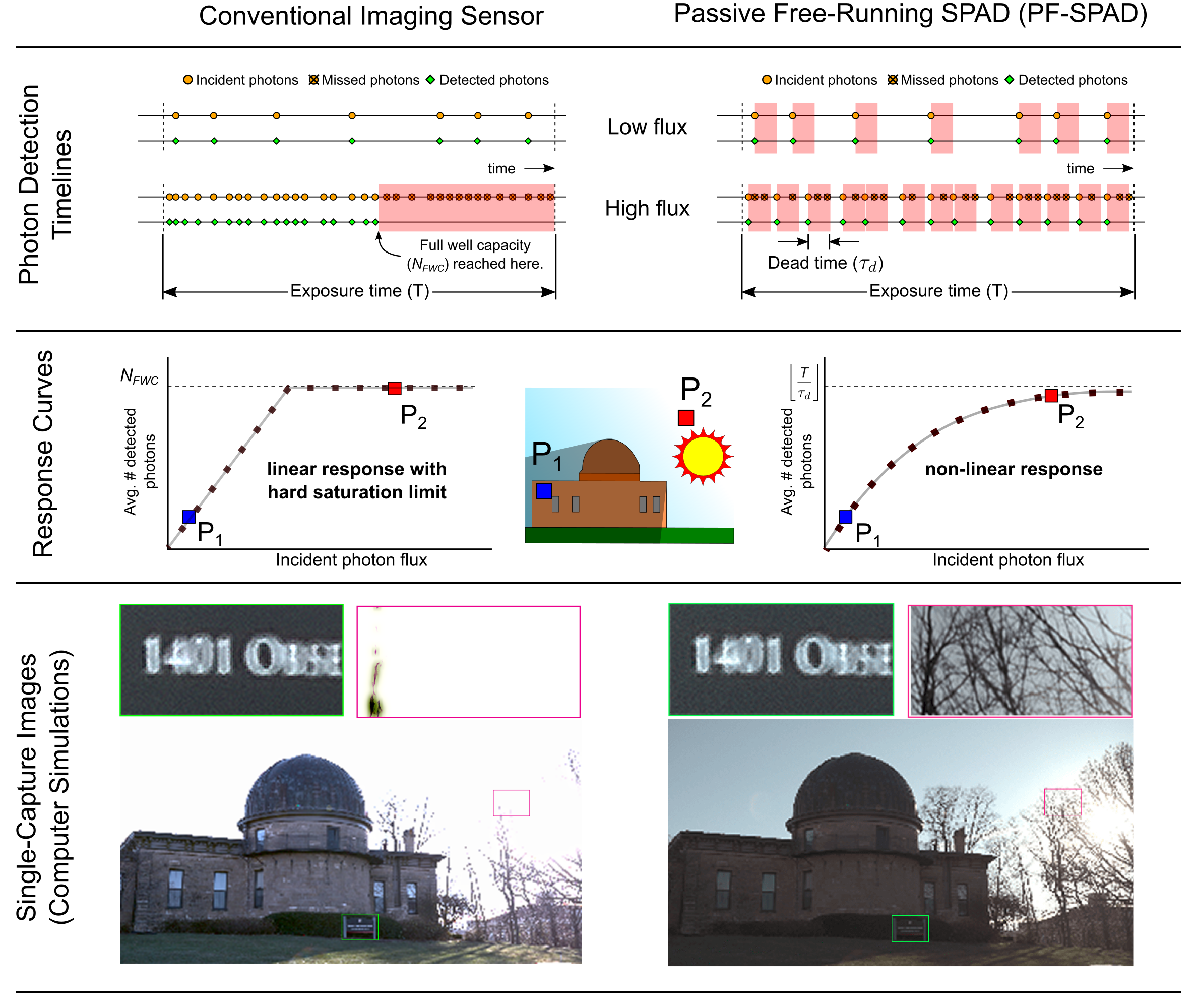}
  \caption{\textbf{Conventional vs. PF-SPAD imaging.} The top row shows
  photon detection timelines at low and high flux levels for the two types of
  sensor pixels. The middle row shows sensor response curves as a function of
  incident photon flux for a fixed exposure time. At high flux, a conventional
  sensor pixel saturates when the full well capacity is reached. A PF-SPAD
  pixel has a non-linear response curve with an asymptotic saturation limit and
  can operate even at extremely high flux levels. The bottom row shows
  simulated single-capture images of an HDR scene with a fixed exposure time of
  $\SI{5}{\milli\second}$ for both types of sensors. The conventional sensor
  has a full well capacity of 33,400. The SPAD has a dead time of
  $\SI{149.7}{\nano\second}$ which corresponds to an asymptotic saturation
  limit equal to 33,400. The hypothetical PF-SPAD array can simultaneously
  capture dark and bright regions of the scene in a single exposure time. The
  PF-SPAD image is for conceptual illustration only; megapixel PF-SPAD arrays
  are currently not available. \label{fig:teaser}}
  \vspace{-0.2in}
\end{figure*}

Single-photon avalanche diodes (SPADs) can count individual photons and capture
their temporal arrival statistics with very high precision~\cite{Cova:96}. Due
to this capability, SPADs are widely used in low light scenarios
\cite{niehorster2016multi,antolovic2017spad,altmann2018review}, LiDAR
\cite{Kirmani58,Shin_2016_naturecomm} and non-line of sight imaging
\cite{Buttafava:15,gariepy2016detection,OToole2018}. In these applications,
SPADs are used in synchronization with an active light source (e.g., a pulsed
laser). In this paper, we propose \emph{passive free-running SPAD} (PF-SPAD)
imaging, where SPADs are used in a \emph{free-running mode}, with the goal of
capturing 2D intensity images of scenes under \emph{passive lighting}, without
an actively controlled light source. Although SPADs have so far been limited to
low flux settings, using the timing statistics of photon arrivals, PF-SPAD
imaging can successfully capture much higher flux levels than previously
thought possible. 

We build a detailed theoretical model and derive a scene brightness estimator for
PF-SPAD imaging that, unlike a conventional sensor pixel, does not suffer from
full well capacity limits~\cite{elgamal2005} and can measure high incident
flux. Therefore, a PF-SPAD remains sensitive to incident light throughout the
exposure time, even under very strong incident flux. This enables imaging
scenes with large brightness variations, from extreme dark to very bright.
Imagine an autonomous car driving out of a dark tunnel on a bright sunny day, or
a robot inspecting critical machine parts made of metal with strong specular
reflections. These scenarios require handling large illumination changes, that
are often beyond the capabilities of conventional sensors.
\smallskip




\noindent {\bf Intriguing Characteristics of PF-SPAD Imaging:} Unlike
conventional sensor pixels that have a linear input-output response (except
past saturation), a PF-SPAD pixel has a non-linear response curve with an
asymptotic saturation limit as illustrated in Figure~\ref{fig:teaser}. After
each photon detection event, the SPAD enters a fixed \emph{dead time} interval
where it cannot detect additional photons.  The non-linear response is a
consequence of the PF-SPAD \emph{adaptively} missing a fraction of the incident
photons as the incident flux increases (see Figure~\ref{fig:teaser} top-right).
Theoretically, a PF-SPAD sensor does not saturate even at
extremely high brightness values. Instead, it reaches a \emph{soft saturation}
limit beyond which it still stays sensitive, albeit with a lower
signal-to-noise ratio (SNR). This soft saturation point is reached considerably
past the saturation limits of conventional sensors, thus, enabling PF-SPADs to
reliably measure high flux values. 

Various noise sources in PF-SPAD imaging also exhibit counter-intuitive
behavior. For example, while in conventional imaging, photon noise increases
monotonically (as square-root) with the incident flux, in PF-SPAD imaging,
the photon noise first increases with incident flux, and then \emph{decreases}
after reaching a maximum value, until eventually, it becomes even lower than the
quantization noise. Quantization noise dominates at very high flux
levels. In contrast, for conventional sensors, quantization noise affects SNR
only at very low flux; and when operating in realistic flux levels, photon
noise dominates other sources of noise. \smallskip




\noindent {\bf Extreme Dynamic Range Imaging with PF-SPADs:} Due to their
ability to measure high flux levels, combined with single-photon
sensitivity, PF-SPADs can simultaneously capture a large range of brightness
values in a single exposure, making them well suited as high dynamic range
(HDR) imaging sensors. We provide theoretical justification for the HDR
capability of PF-SPAD imaging by modeling its photon detection statistics. We
build a hardware prototype and demonstrate single-exposure imaging of scenes
with an extreme dynamic range of $10^6\!\!:\!\!1$, over 2 orders of magnitude higher
than conventional sensors. We envision that the proposed approach and analysis
will expand the applicability of SPADs as \emph{general-purpose,
all-lighting-condition, passive} imaging sensors, not limited to specialized
applications involving low flux conditions or active illumination, and play a
key role in applications that witness extreme variations in flux levels,
including astronomy, microscopy, photography, and computer vision systems. 
\smallskip

\noindent{\bf Scope and Limitations:}
The goal of this paper is to present the concept of \emph{adaptive temporal
binning} for passive flux sensing and related theoretical analysis using a
single-pixel PF-SPAD implementation. Current SPAD technology is still in
a nascent stage, not mature enough to replace conventional CCD and CMOS image
sensors.  Megapixel PF-SPAD arrays have not been realized yet. Various
technical design challenges that must be resolved to enable high resolution
PF-SPAD arrays are beyond the scope of this paper.



\section{Related Work}
\noindent{\bf HDR Imaging using Conventional Sensors:} The key idea behind HDR
imaging with digital CMOS or CCD sensors is similar to combination printing
\cite{robinson1860printing} --- capture more light from darker parts of the
scene to mitigate sensor noise and less light from brighter parts of the scene
to avoid saturation. A widely used computational method called exposure
bracketing \cite{debevec2008recovering,Gupta_2013} captures multiple images of the scene
using different exposure times and blends the pixel values to generate an HDR
image.Exposure bracketing algorithms can be adapted to the
PF-SPAD image formation model to further increase their dynamic range.

\noindent{\bf Hardware Modifications to Conventional Sensors:} Spatially
varying exposure technique modulates the amount of light reaching the sensor
pixels using fixed \cite{Nayar_2000} or adaptive \cite{Nayar_2003} light
absorbing neutral density filters. Another method \cite{tocci2011versatile}
involves the use of beam-splitters to relay the scene onto multiple imaging
sensors with different exposure settings. In contrast, our method can provide
improved dynamic range without having to trade off spatial resolution.

\noindent{\bf Sensors with Non-Linear Response:}
Logarithmic image sensors \cite{Kavadias_2000} use additional hardware in each
pixel that applies logarithmic non-linearity to obtain dynamic range
compression. Quanta image sensors (QIS) obtain logarithmic dynamic range
compression by exploiting fine-grained (sub-diffraction-limit) spatial
statistics, through spatial oversampling
\cite{Feng_Yang_2012,Fossum_2016,Dutton_2016}. We take a different approach of
treating a SPAD as an \emph{adaptive temporal binary sensor} which subdivides
the total exposure time into random non-equispaced time bins at least as long
as the dead time of the SPAD.  Experimental results in recent work
\cite{charbon2018} have shown the potential of this method for improved dynamic
range over the QIS approach. Here we provide a comprehensive theoretical
justification by deriving the SNR from first principles and also show simulated
and experimental imaging results demonstrating dynamic range improvements of
over two orders of magnitude.

\section{Passive Imaging with a Free-Running SPAD}
In this section we present an image formation model for a PF-SPAD and derive a
photon flux estimator that relies on inter-photon detection times and
photon counts. This provides formal justification for the notion of adaptive
photon rejection and the asymptotic response curve of a PF-SPAD.

Each PF-SPAD pixel passively measures the photon flux from a scene point by
detecting incident photons over a fixed exposure time. The time intervals
between consecutive incident photons vary randomly according to a Poisson
process \cite{Hasinoff2014}. If the difference in the arrival times of two
consecutive photons is less than the SPAD dead time, the later photon is not
detected. The free-running operating mode means that the PF-SPAD pixel is ready
to capture the next available photon as soon as the dead time interval from the
previous photon detection event elapses\footnote{In contrast, conventionally,
SPADs are triggered at fixed intervals, for example, synchronized with a laser
pulse in a LiDAR application, and the SPAD detects at most one photon for each
laser pulse.}. In this free-running, passive-capture mode the PF-SPAD pixel
acts as a temporal binary sensor that divides the total exposure time into
random, non-uniformly spaced time intervals, each at least as long as the dead
time. As shown in Figure~\ref{fig:teaser}, the PF-SPAD pixel detects at most
one photon within each interval; additional incident photons during the dead
time interval are not detected. The same figure also shows that as the average
number of photons incident on a SPAD increases, the fraction of the number of
detected photons decreases.
\smallskip

\noindent{\bf PF-SPAD Image Formation Model:}
Suppose the PF-SPAD pixel is exposed to a constant photon flux of $\Phi$
photons per unit time over a fixed exposure time $T$. Let $N_T$ denote the
total number of photons detected in time $T$, and
$\{X_1,X_2,\ldots,X_{N_T-1}\}$ denote the inter-detection time intervals. We
define the \emph{average time of darkness} as $\bar X =
\frac{1}{N_T-1}\sum_{i=1}^{N_T-1}{X_i}$.  Intuitively, a larger incident flux
should correspond to a lower average time of darkness, and vice versa. Based on
this intuition, we derive the following estimator of the incident flux as a
function of $\bar X$ (see \nolink{\ref{sec:suppl_note_spad_image_formation}}
for derivation): 
\smallskip
\begin{equation} \hat\Phi = \frac{1}{q \left(\bar X - \tau_d \right)},
  \label{eq:spad_flux_from_mean_interarrival_time}
\end{equation}
\smallskip
where $\hat\Phi$ denotes the estimated photon flux, $0<q<1$ is the photon
detection probability of the SPAD pixel, and $\tau_d$ is the dead time. Note
that since $X_i \geq \tau_d \,\, \forall \,i$, the estimator in
Equation~(\ref{eq:spad_flux_from_mean_interarrival_time}) is positive and
finite.  In a practical implementation, it is often more efficient to use fast
counting circuits that only provide a count of the total number of SPAD
detection events in the exposure time interval, instead of storing timestamps
for individual detection events. In this case, the average time of darkness can
be approximated as $\bar X \approx T/N_T$. The flux estimator that uses only
photon counts is given by:

\begin{equation}
\underbrace{\boxed{
  \hat \Phi = \,\frac{N_T}{q \left(T - N_T\,\tau_d \right)}
}}_\text{\textbf{PF-SPAD Flux Estimator}}.
  \label{eq:spad_flux_from_counts}
\end{equation}

\noindent{\bf Interpreting the PF-SPAD Flux Estimator:}
The photon flux estimator in Equation~(\ref{eq:spad_flux_from_counts}) is a
function of the number of photons detected by a dead time-limited SPAD pixel
and is valid at all incident flux levels.  The image formation procedure
applies this inverse non-linear mapping to the photon counts from each PF-SPAD
pixel to recover flux values, even for bright parts of the scene. The
relationship between the estimated flux, $\hat\Phi$, and the number of photons
detected, $N_T$, is non-linear, and is similar to the well-known
non-paralyzable detector model used to describe certain radioactive particle
detectors \cite{muller1973dead,grimmett_probbook_2001}.

To obtain further insight into the non-linear behaviour of a SPAD pixel in
the free-running mode, it is
instructive to analyze the average number of detected photons as a function of
$\Phi$ for a fixed $T$. Using the theory of renewal processes
\cite{grimmett_probbook_2001} we can show that:
\begin{equation}
  \mathbf E[N_T] = \frac{q\Phi T}{1+q\Phi \tau_d}\,. \label{eq:spad_avg_counts}
\end{equation}
This non-linear SPAD response curve is shown in Figure~\ref{fig:teaser}.  The
non-linear behavior is a consequence of the ability of a SPAD to perform
adaptive photon rejection during the exposure time. The shape of the response
curve is similar to a gamma-correction or tone-mapping curve used for
displaying an HDR image. As a result, the SPAD response curve provides dynamic
range compression, \emph{gratis}, with no additional hardware modifications.
The key observation about Equation~(\ref{eq:spad_avg_counts}) is that it has an
asymptotic saturation limit given by $\lim_{\Phi \rightarrow \infty}\mathbf
E[N_T] = T/ \tau_d$. Therefore, in theory, the photon counts never saturate
because this asymptotic limit can only be achieved with an infinitely bright
light source.  In practice, as we discuss in the following sections, due to the
inherent quantized nature of photon counts, the estimator in
Equation~(\ref{eq:spad_flux_from_counts}) suffers from a soft saturation
phenomenon at high flux levels and limits the SNR.

\section{Peculiar Noise Characteristics of PF-SPADs}
In this section, we list various noise sources that affect a PF-SPAD pixel,
derive mathematical expressions for the bias and variance they introduce in the
total photon counts, and provide intuition on their surprising,
counter-intuitive characteristics as compared to a conventional pixel.
Ultimately, the flux estimation performance limits will be determined by the
cumulative effect of these sources of noise as a function of the incident
photon flux.
\smallskip

\begin{figure}[!ht]
  \centering \includegraphics[width=0.95\columnwidth]{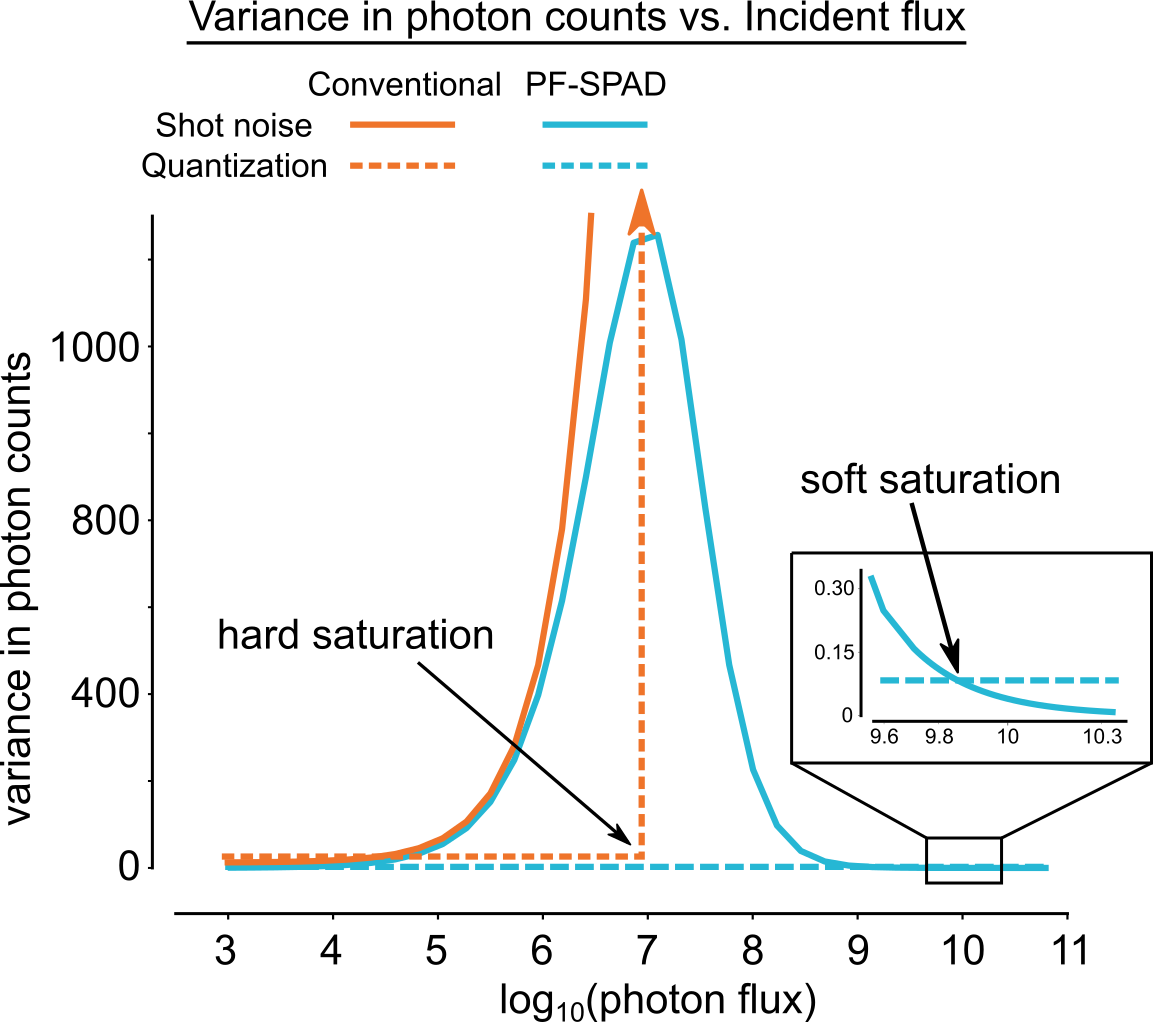}
  \caption{\textbf{Effect of various sources of noise on variance of PF-SPAD
  photon counts.} For a PF-SPAD pixel, the variance in photon counts due to
  quantization remains constant at all flux levels. The variance due to shot
  noise first increases and then decreases with increasing incident flux. At
  the soft saturation point, quantization exceeds shot noise variance.
  For a conventional pixel, quantization noise remains small and constant until
  the full well capacity is reached, where it jumps to infinity. Shot noise
  variance increases monotonically with incident flux.
  \label{fig:spad_sources_of_noise}}
  \vspace{-0.1in}
\end{figure}

\noindent{\bf Shot Noise:} For a conventional image sensor, due to
Poisson distribution of photon arrivals, the variance of shot noise is
proportional to the incident photon flux~\cite{Hasinoff2014}, as shown in
Figure~\ref{fig:spad_sources_of_noise}. A PF-SPAD, however, adaptively rejects
a fraction of the incident photons during the dead time. Therefore, although
the incident photons follow Poisson statistics, the photon counts (number of
\emph{detected} photons) do not. We define shot noise for PF-SPADs as the
variance in the detected number of photon counts. This is approximately given
by (see \nolink{\ref{sec:suppl_note_spad_approx_snr}}):
\begin{equation}
  \mathrm{Var}[N_T] = \frac{q \, \Phi \, T}{(1 + q \, \Phi \, \tau_d)^3}. \label{eq:spad_photon_variance}
\end{equation}
As shown in Figure~\ref{fig:spad_sources_of_noise}, the variance first
increases as a function of incident flux, reaches a maximum and then decreases
at very high flux levels. This peculiar behavior can be understood intuitively
from the PF-SPAD photon detection timelines in Figure~\ref{fig:teaser} and
observing how the dead time intervals are spread within the exposure time. At
low flux, when $\Phi \ll 1/\tau_d$, the dead time windows, on average, have
large intervening time gaps. So the detected photon count statistics behave
approximately like a conventional image sensor with Poisson statistics:
$\mathrm{Var}[N_T] \approx q \Phi T$. This explains the monotonically
increasing trend in variance at low flux.  However, for large incident flux
$\Phi \gg 1/\tau_d$ the time of darkness between consecutive dead time windows
becomes sufficiently small that the PF-SPAD detects a photon soon after the
preceding dead time interval expires.  This causes a \emph{decrease} in
randomness which manifests as a monotonically decreasing photon count variance.
In theory, as $\Phi \rightarrow \infty$ the process becomes deterministic with
zero variance: the PF-SPAD detects exactly one photon per dead time window.
\smallskip

\noindent\textbf{Quantization Noise and Saturation:} For a PF-SPAD, since the
photon counts are always integer valued, the source of quantization noise is
inherent in the measurement process. As a first order approximation, this can
be modeled as being uniformly distributed in the interval $[0,1]$ which has a
variance of $\nicefrac{1}{12}$ for all incident flux levels.\footnote{For exact
theoretical analysis refer to \nolink{\ref{sec:suppl_note_spad_exact_snr}}.}
A surprising consequence of the monotonically decreasing behavior of PF-SPAD
shot noise is that at sufficiently high photon flux, quantization noise
exceeds shot noise and becomes the \emph{dominant} source of noise. This
shown in Figure~\ref{fig:spad_sources_of_noise}~(zoomed inset). We refer to
this phenomenon as \mbox{\emph{soft saturation}}, and discuss this in more
detail in the next section.

In contrast, for a conventional imaging sensor, quantization noise is often
ignored at high incident flux levels because state of the art CMOS and CCD
sensors have analog-to-digital conversion (ADC) with sufficient bit depths.
However, these sensors suffer from full well capacity limits beyond which they
can no longer detect incident photons. As shown in
Figure~\ref{fig:spad_sources_of_noise}, we incorporate this hard saturation
limit into quantization noise by allowing the quantization variance to jump to
infinity when the full well capacity is reached. \smallskip



\noindent {\bf Dark Count and Afterpulsing Noise:} Dark counts are spurious
counts caused by thermally generated electrons and can be modeled as a Poisson
process with rate $\Phi_\mathrm{dark}$, independent of the true photon
arrivals. Afterpulsing noise refers to spurious counts caused due to charged
carriers that remain trapped in the SPAD from preceding photon detections. In
most modern SPAD detectors dark counts and afterpulsing effects are usually
negligible and can be ignored. \smallskip

\noindent {\bf Effect of Noise on Scene Brightness Estimation:} Since the
output of a conventional sensor pixel is linear in the incident brightness, the
variance in estimated brightness is simply equal (up to a constant scaling
factor) to the noise variance. This is not the case for a PF-SPAD pixel due to
its non-linear response curve --- the variance in photon counts due to
different sources of noise must be converted to a variance in brightness
estimates, by accounting for the non-linear dependence of $\hat\Phi$ on $N_T$
in Equation~(\ref{eq:spad_flux_from_counts}). This raises a natural question:
Given the various noise sources that affect the photon counts obtained from a
PF-SPAD pixel, how reliable is the estimated scene brightness?


\section{Extreme Dynamic Range of PF-SPADs}
The various sources of noise in a PF-SPAD pixel described in the previous
section cause the estimated photon flux $\hat\Phi$ to deviate from the true
value $\Phi$. In this section we derive mathematical expressions for the bias
and variance introduced by these different sources of noise in the PF-SPAD flux
estimate. The cumulative effect of these errors is captured in the
root-mean-squared error (RMSE) metric:
\smallskip
\begin{equation*}
\mathsf{RMSE}(\hat \Phi) = \sqrt{\mathbf{E}[(\hat\Phi-\Phi)^{2}]},
\end{equation*}
where the expectation operation averages over all the sources of noise in the
SPAD pixel. Using the bias-variance decomposition, the RMSE of the PF-SPAD flux
estimator can be decomposed as a sum of flux estimation errors from the
different sources of noise:

\begin{figure*}[!ht]
  \centering \includegraphics[width=0.94\textwidth]{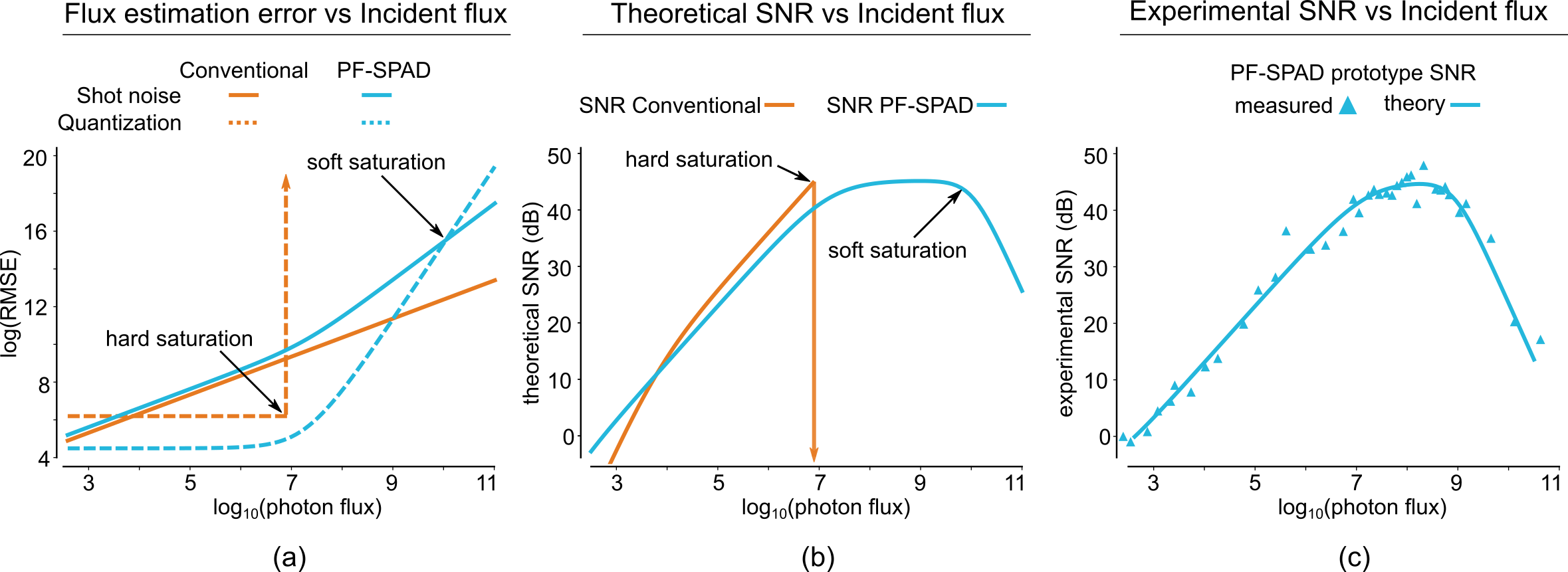}
  \caption{\textbf{Signal-to-noise ratio of a PF-SPAD pixel.} (a) A PF-SPAD
  pixel suffers from quantization noise, which results in flux estimation error
  that increases as a function of incident flux. Beyond a flux level denoted as
  ``soft saturation,'' quantization becomes the dominant noise source
  overtaking shot noise. In contrast, for conventional sensors, quantization
  and read noise remain constant while shot noise increases with incident flux.
  (b) Unlike a conventional sensor, a PF-SPAD sensor does not suffer from a
  hard saturation limit. A soft saturation response leads to a graceful drop in
  SNR at high photon flux, leading to a high dynamic range. (c) An experimental
  SNR plot obtained from a hardware prototype consisting of a
  $\SI{25}{\micro\meter}$ PF-SPAD pixel with a $149.7\pm\SI{6}{\nano\second}$
  dead time and $\SI{5}{\milli\second}$ exposure time.\label{fig:snr}}
  \vspace{-0.1in}
\end{figure*}


\begin{equation}
\mathsf{RMSE}(\hat\Phi)\! =\! \sqrt{(\Phi_\mathrm{dark}\! +\! B_\mathrm{ap})^2 +\!
  V_\text{shot}\! +\! V_\mathrm{quantization}}\,. \label{eq:spad_rmse}
\end{equation}
The variance in the estimated flux due to shot noise
(Equation~(\ref{eq:spad_photon_variance})) is given by:
\begin{equation}
V_\mathrm{shot} = \frac{\Phi(1+q\Phi\tau_{d})}{qT}.\label{eq:V_shot}
\end{equation}
The variance in estimated flux due to quantization is:
\begin{equation}
V_\mathrm{quantization} = \frac{(1+q\Phi\tau_{d})^{4}}{12q^{2}T^{2}}.
\label{eq:V_quantization}
\end{equation}
The dark
count bias $\Phi_\mathrm{dark}$ depends on the operating temperature. Finally,
the afterpulsing bias $B_\mathrm{ap}$ can be expressed in terms of the
afterpulsing probability $p_\mathrm{ap}$:
\begin{equation}
B_\mathrm{ap} = p_\mathrm{ap} \, q\Phi \, (1 +
\Phi\tau_{d})e^{-q\Phi\tau_{d}}. \label{eq:B_ap}
\end{equation}
See \nolink{\ref{sec:suppl_note_spad_approx_snr}} and
\nolink{\ref{sec:suppl_note_spad_exact_snr}} for detailed derivations of
Equations~(\ref{eq:V_shot}--\ref{eq:B_ap}).

Figure~\ref{fig:snr}(a) shows the flux estimation errors introduced by the
various noise sources as a function of the incident flux levels for a
conventional and a PF-SPAD pixel.\footnote{The effects of dark counts and
afterpulsing noise are usually negligible and are discussed in
\nolink{\ref{sec:suppl_note_spad_sources_of_noise}} and shown in
Supplementary Figure~\ref{fig:spad_sources_of_noise_flux_estimate}.}
The performance of the PF-SPAD flux estimator can be expressed in terms of its
SNR, formally defined as the ratio of the true photon flux to the RMSE of the
estimated flux \cite{Feng_Yang_2012}:
\begin{equation}
\mathsf{SNR}(\Phi)=20\log_{10}\left(\frac{\Phi}{
  \mathsf{RMSE}(\hat\Phi)}\right).\label{eq:snr_general_formula}
\end{equation}
By substituting the expressions for various noise sources from
Equations~(\ref{eq:spad_rmse}-\ref{eq:V_quantization}) into
Equation~(\ref{eq:snr_general_formula}), we get an expression for the SNR of
the SPAD-based flux estimator shown in
Equation~(\ref{eq:spad_snr_ap_corrected}). Figure~\ref{fig:snr}(b) shows the
theoretical SNR as a function of incident flux for the PF-SPAD flux estimator,
and a conventional sensor. A conventional sensor suffers from an abrupt drop in
SNR due to hard saturation (see \nolink{\ref{sec:suppl_note_ccd}}).  In
contrast, the SNR achieved by a SPAD sensor degrades gracefully, even beyond
the soft saturation point.  \smallskip

\noindent{\bf The Soft Saturation Phenomenon:} It is particularly instructive
to observe the behavior of quantization noise for the SPAD pixel. Although the
quantization noise in the detected photon counts remains small and constant at
all flux levels, the variance in the estimated flux due to quantization
increases monotonically with incident flux. This is due to the non-linear
nature of the estimator in Equation~(\ref{eq:spad_flux_from_counts}). At high
incident flux levels, a single additional detected photon maps to a large range
of estimated flux values, resulting in large errors in estimated flux. We call
this phenomenon \emph{soft saturation}. Beyond the soft saturation flux level,
quantization dominates all other noise sources, including shot noise. The soft
saturation limit, however, is reached at considerably higher flux levels as
compared to the hard saturation limit of conventional sensors, thus, enabling
PF-SPADs to reliably estimate very high flux levels. 
\smallskip

\begin{figure*}[!ht]
\hrulefill
\normalsize
\setcounter{MYtempeqncnt}{\value{equation}}
\setcounter{equation}{9}
\begin{equation}
\boxed{ \mathsf{SNR}(\Phi)\! =\! - 10\log_{10} \left[
  \left(
  \frac{\Phi_\mathrm{dark}}{\Phi}\! +\! q(1\! +\! \Phi\tau_{d})p_\mathrm{ap}e^{-q\Phi\tau_{d}}
  \right)^{2}\! + \!
  \frac{(1+q\Phi\tau_{d})}{q \Phi T}\! +\! \frac{(1+q\Phi\tau_{d})^{4}}{12q^{2}\Phi^2 T^{2}}
  \right]}\,. \label{eq:spad_snr_ap_corrected}
\end{equation}
\setcounter{equation}{4}
\hrulefill
\vspace*{0pt}
\vspace{-0.1in}
\end{figure*}

\noindent{\bf Effect of Varying Exposure Time:}
For conventional imaging sensors, increasing the exposure time causes the
sensor pixel to saturate at a lower value of the incident flux level. This is
equivalent to a horizontal translation of the conventional sensor's SNR curve
in Figure~\ref{fig:snr}(b). This does not affect its dynamic range.  However,
for a PF-SPAD pixel, the asymptotic saturation limit increases linearly with
the exposure time, hence increasing the SNR at all flux levels. This leads to a
remarkable behavior of increasing the dynamic range of a PF-SPAD pixel with
increasing exposure time. See~\nolink{\ref{sec:suppl_note_varying_exp_time}}
and Supplementary Figure~\nolink{\ref{fig:snr_effect_of_parameters}}.
\smallskip

\noindent{\bf Simulated Megapixel PF-SPAD Imaging System:}
Figure~\ref{fig:teaser} (bottom row) shows simulated images for a conventional
megapixel image sensor array and a \emph{hypothetical} megapixel PF-SPAD array.
The ground truth photon flux image was obtained from an exposure bracketed HDR
image captured using a Canon EOS Rebel T5 DSLR camera with 10 stops rescaled to
cover a dynamic range of $10^6:1.$ An exposure time of
$T=\SI{5}{\milli\second}$ was used to simulate both images. For fair comparison, the
SPAD dead time was set to \SI{149.7}{\nano\second}, which corresponds to an
asymptotic saturation limit of $\nicefrac{T}{\tau_d} = \num{34000}$, equal to the
conventional sensor full well capacity. The quantum efficiencies of the
conventional sensor and PF-SPAD were set to 90\% and 40\%. Observe that the
PF-SPAD can simultaneously capture details in the dark regions of the scene
(e.g. the text in the shadow) and bright regions in the sun-lit sky. The
conventional sensor array exhibits saturation artifacts in the bright regions
of the scene. (See \nolink{\ref{sec:suppl_simulation_model}}).

The human eye has a unique ability to adapt to a wide range of brightness
levels ranging from a bright sunny day down to single photon levels
\cite{Blackwell_1946,Tinsley_2016}. Conventional sensors cannot simultaneously
reliably capture very dark and very bright regions in many natural scenes.
In contrast, a PF-SPAD can simultaneously image dark and bright regions of the
scene in a single exposure. Additional simulation results are shown in
Supplementary
Figures~\nolink{\ref{fig:simulated_observatory}--\ref{fig:simulated_indoor_lcpr}}.

\section{Experimental Results}
\paragraph{SNR and Dynamic Range of a Single-Pixel PF-SPAD:}
Figure~\ref{fig:snr}(c) shows experimental SNR
measurements using our prototype single-pixel SPAD sensor together with the SNR
predicted by our theoretical model.
Our hardware prototype has an additional $\SI{6}{\nano\second}$ jitter
introduced by the digital electronics that control the dead time window
duration. This is not included in the SNR curve of Figure~\ref{fig:snr}(b) but
is accounted for in the theoretical SNR curve shown in Figure~\ref{fig:snr}(c).
See \nolink{\ref{sec:suppl_note_spad_dead_time_unc}} for details. We define
dynamic range as the ratio of largest to smallest photon flux values that can
be measured above a specified minimum SNR.  Assuming a minimum acceptable SNR
of 30 dB, the SPAD pixel achieves a dynamic range improvement of \emph{over 2
orders of magnitude} compared to a conventional sensor. \smallskip

\begin{figure}[!ht]
  \centering \includegraphics[width=0.85\columnwidth]{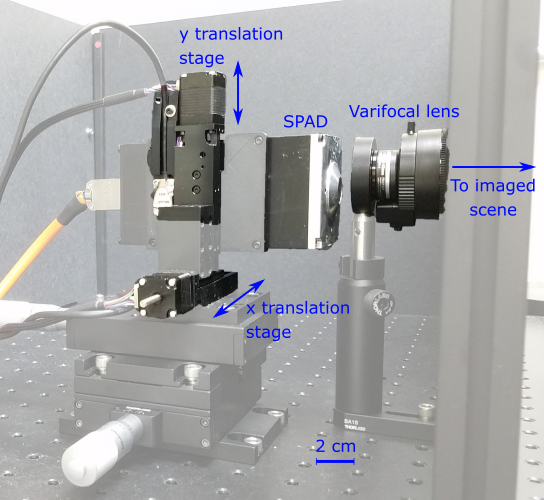}
  \caption{{\bf Experimental single-pixel PF-SPAD imaging system.} 
  A free-running SPAD is mounted on two translation stages to raster-scan the
  image plane. There is no active light source---the PF-SPAD passively measures
  ambient light in the scene. Photon counts are captured using a single-photon
  counting module (not shown) operated without a synchronization signal.
    \label{fig:hardware_setup}}
    \vspace{-0.04in}
\end{figure}

\noindent\textbf{Point-Scanning Setup:}
The imaging setup shown in Figure~\ref{fig:hardware_setup} consists of a SPAD
module mounted on a pair of micro-translation stages (VT-21L Micronix USA)
to raster-scan the image plane of a variable focal length lens (Fujifilm
DV3.4x3.8SA-1). Photon counts were recorded using a single-photon counting
module (PicoQuant HydraHarp 400), with the SPAD in the free-running mode. A
monochrome machine vision camera (FLIR GS3-U3-23S6M-C) was used for qualitative
comparisons with the images acquired using the SPAD setup. The machine vision
camera uses the same variable focal length lens with identical field of view as
the scene imaged by the SPAD point-scanning setup. This ensures a comparable
effective incident flux on a per-pixel basis for both the SPAD and the machine
vision camera. The sensor pixel parameters are identical to those used in
simulations. Images captured with the machine vision camera were downsampled to
match the resolution of the raster-scanned PF-SPAD images.
\smallskip


\begin{figure*}[!ht]
  \centering \includegraphics[width=1.0\textwidth]{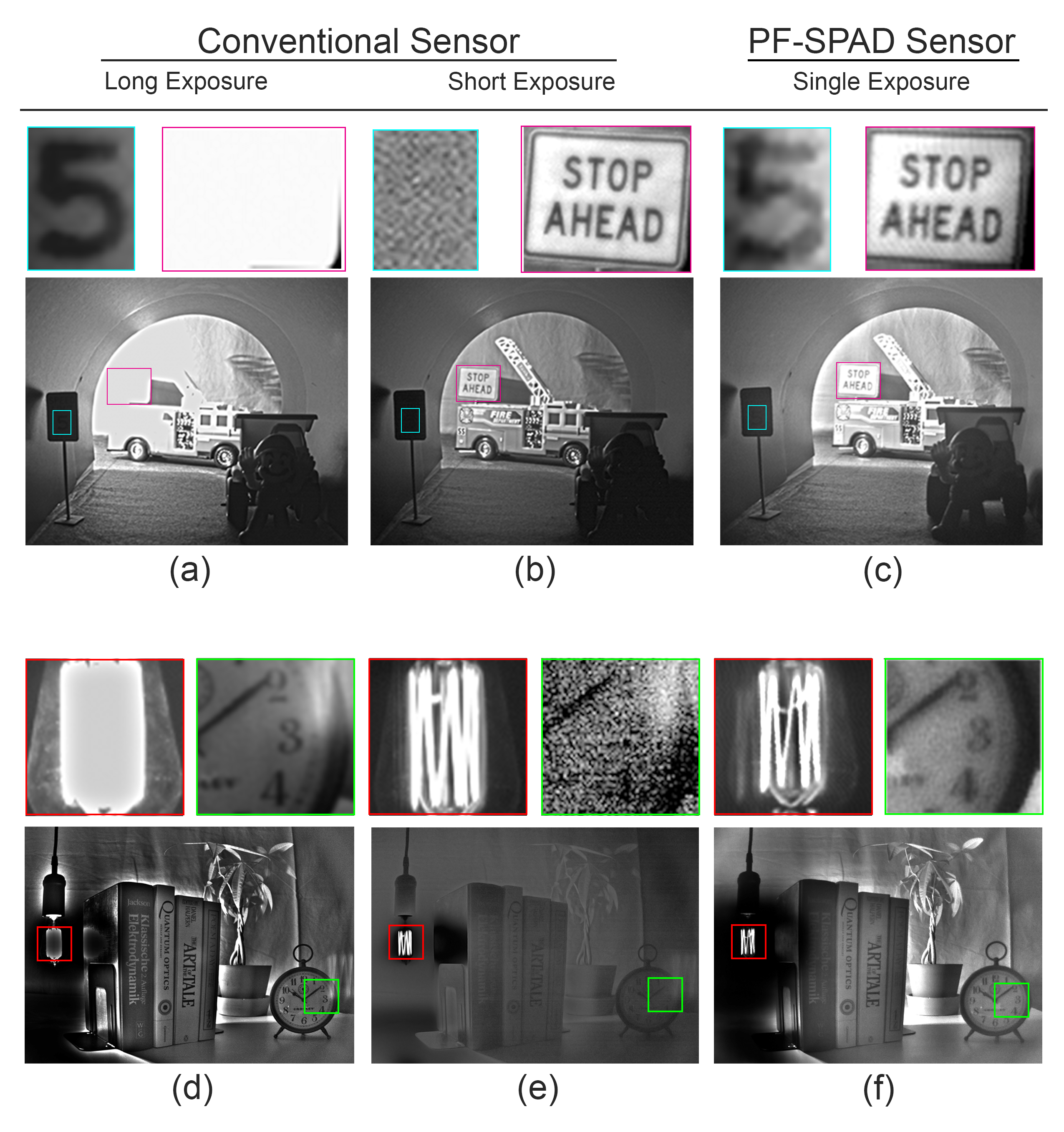}
  \caption{
  \textbf{Experimental comparison of the dynamic range of a CMOS
  camera and PF-SPAD imaging.} The two imaged scenes have a wide range of
  brightness values (1,000,000:1), considerably beyond the dynamic range of
  conventional sensors. (a, d) Images captured using a 12-bit CMOS machine
  vision camera with a long exposure time of $\SI{5}{\milli\second}$. Bright
  regions appear saturated. (b, e) Images of the same scenes with a short
  exposure time of $\SI{0.5}{\milli\second}$. Darker regions appear grainy and
  severely underexposed, making it challenging to read the text on the signs
  and the numbers on the alarm clock. (c, f) PF-SPAD images of the same scenes
  captured using a single $\SI{5}{\milli\second}$ exposure per pixel. Our
  hardware prototype captures the full range of brightness levels in the scenes
  in a single shot.  The text is visible in both bright and dark regions of the
  scene, and details in regions of high flux, such as the filament of the bulb,
  can be recovered. For fair comparison, the main images were tone-mapped
  using the same tone-mapping algorithm.}
  \label{fig:pointscan}
\end{figure*}



\noindent\textbf{Extreme HDR:}
Results of single-shot HDR images from our raster-scanning PF-SPAD prototype
are shown in Figure~\ref{fig:pointscan} and
Supplementary~Figure~\ref{fig:artistic_scene_2}, for different scenes spanning
a wide dynamic range ($\geq 10^6:1$) of brightness values.
To reliably visualize the wide range of brightnesses in these scenes,
three different tone-mapping algorithms were used to tone-map the main figures,
the dark zoomed insets and the bright zoomed insets, respectively. The machine vision
camera fails to capture bright text outside the tunnel
(Fig.~\ref{fig:pointscan}(a)) and dark text in the tunnel
(Fig.~\ref{fig:pointscan}(b)) in a single exposure interval. The PF-SPAD
successfully captures the entire dynamic range (Fig.~\ref{fig:pointscan}(c)).
In Fig.~\ref{fig:pointscan}(f), the PF-SPAD even captures the bright filament
of an incandescent bulb simultaneously with dark text in the shadow. 
The halo artifacts in Figure~\ref{fig:pointscan}(d--f) are due to a local
adaptation-based non-invertible tone-map that was used to simultaneously
visualize the bright filament and the dark text.  This ability of the PF-SPAD
flux estimator to capture a wide range of flux from very low to high in a
single capture can have implications in many
applications~\cite{Marois_2008,Brady_2012,Vinegoni_HDR_Microscopy_2016,WaymoPatent_2017}.
that require extreme dynamic range.


\section{Discussion}
\noindent\textbf{Quanta Image Sensor:}
An alternative realization \cite{Itzler_Patent_2017} of a SPAD-based imaging
sensor divides the total exposure time $T$ into uniformly spaced intervals of
duration $\tau_b \geq \tau_d$. This ``uniform-binning'' method leads to a
different image formation model which is known in literature as the oversampled
binary image sensor \cite{Feng_Yang_2012} or quanta image sensor (QIS)
\cite{Fossum_2016,Dutton_2016}. In \nolink{\ref{sec:suppl_note_qis_snr}}, we show that 
in theory, this uniform-binning implementation has a smaller dynamic range as
compared to a PF-SPAD that allows the dead time windows to shift adaptively
\cite{charbon2018}. Note, however, that state of the art QIS
technology provides much higher resolution and fill factor with high quantum
efficiencies, and lower read noise than current SPAD arrays.
\smallskip

\noindent\textbf{Limitations and Future Outlook:}
Our proof-of-concept imaging system uses a SPAD that is not optimized for
operating in the free-running mode. The duration of the dead time window,
which is a crucial parameter in our flux estimator, is not stable in current
SPAD implementations (such as silicon photo-multipliers) as it is not crucial
for active time-of-flight applications. Various research and engineering
challenges must be met to realize a high resolution SPAD-based passive image
sensor. State of the art SPAD pixel arrays that are commercially available
today consist of thousands of pixels with row or column multiplexed readout
capabilities and do not support fully parallel readout. Current SPAD arrays
also have very low fill factors due to the need of integrating counting and
storage electronics within each pixel \cite{small_tdc_2017,gyongy_2018}. Our
method and results make a case for developing high resolution fabrication and
3D stacking techniques that will enable high fill-factor SPAD arrays, which can
be used as general purpose, passive sensors for applications requiring extreme
dynamic range imaging.
\smallskip


%


\clearpage
\onecolumn
\renewcommand{\figurename}{Supplementary Figure}
\renewcommand{\thesection}{Supplementary Note \arabic{section}}
\renewcommand{\theequation}{S\arabic{equation}}
\setcounter{figure}{0}
\setcounter{algorithm}{0}
\setcounter{section}{0}
\setcounter{equation}{0}
\setcounter{page}{1}
\renewcommand*{\thefootnote}{$\dagger$}

\begin{center}
\huge Supplementary Document for\\[0.5cm]
\huge ``\mytitle'' \\[0.7cm]
\normalsize Atul Ingle, Andreas Velten$^\dagger$, Mohit Gupta\footnote{Equal contribution.}\\[0.5cm]
Correspondence to: ingle@uwalumni.com\\
\end{center}

\renewcommand*{\thefootnote}{\arabic{footnote}}

\section{Image Formation Model and Flux Estimator for a PF-SPAD
  Pixel\label{sec:suppl_note_spad_image_formation}}
A PF-SPAD sensor pixel and a time-correlated photon counting module are used
to obtain total photon counts over a fixed exposure time together with
picosecond resolution measurements of the time elapsed between successive
photon detection events. We will assume that the PF-SPAD pixel is exposed to a
true photon flux of $\Phi$ photons/second for an exposure time of $T$ seconds
and it records $N_T$ photons in exposure time interval $(0,T]$. For
mathematical convenience, we assume that the exposure interval starts with a
photon detection event at time $t=0$.

Photons arrive at the SPAD according to a Poisson process.
Accounting for an imperfect photon detection efficiency of $0 < q <1$,
the time between consecutive incident photons follows an
exponential distribution with rate $q\Phi.$ After each detection event,
the SPAD enters a dead time window of duration $\tau_d$. Due to the memoryless
property of Poisson processes [\ref{sr1}], the time interval between the
end of a dead time window and the next photon arrival is also exponentially
distributed and has the same rate $q\Phi$ as the incident Poisson process.
Let $X_1$ be the time of the first photon detection after $t=0$ and $X_n$
be the time between the $n-1^{st}$ and $n^{th}$ detection event for
$n \geq 2$. Then the inter-detection time duration $X_n$ follows a
\emph{shifted} exponential distribution given by:

\begin{equation}
X_{n}\stackrel{iid}{\sim}f_{X_{n}}(t)=\begin{cases}
q\Phi e^{-q\Phi(t-\tau_{d})} & \mbox{for }t\geq\tau_{d}\\
0 & \mbox{otherwise.}
\end{cases} \label{eq:interarrival_distr}
\end{equation}

This provides a probabilistic model of the photon inter-detection times. We now
derive a flux estimator from a sequence of observed inter-detection times captured
by a PF-SPAD pixel.

\subsection*{Estimating Flux from Inter-Detection Time
Intervals}
The log-likelihood function for the observed inter-detection times is given by

\begin{align}
  \log l(q\Phi; X_1,\ldots,X_{N_T}) &= \log \left( \prod_{n=1}^{N_T} q\Phi\,e^{-q\Phi(X_n-\tau_{d})} \right) \nonumber \\
  &= - q\Phi \left( \sum_{n=1}^{N_T}X_n - \tau_d N_T \right) + N_T \log q\Phi \nonumber \\
  &= - q\Phi\, N_T\,\left( \bar X - \tau_d \right) + N_T \log q\Phi \label{eq:loglik}
\end{align}
where $\bar X := \frac{1}{N_T} \sum_{n=1}^{N_T} X_n$ is the mean time between
photon detection events.  The maximum likelihood estimate $\hat \Phi$ of the
true photon flux is computed by setting the derivative of Equation~(\ref{eq:loglik}) to
zero:

$$
\frac{N_T}{q\hat \Phi} - N_T (\bar X - \tau_d) = 0
$$
which implies

\begin{equation}
  \hat \Phi = \frac{1}{q} \frac{1}{\bar X - \tau_d}. \label{eq:flux_from_interarrival_times}
\end{equation}

\clearpage
\section{Approximate Closed Form Formula for SNR of a PF-SPAD
  pixel\label{sec:suppl_note_spad_approx_snr}}
We first derive an approximate formula for the SNR of a PF-SPAD pixel using a continuous
Gaussian distribution approximation for the number of counts $N_T$. The effective incident
photon flux for a quantum efficiency of $0<q<1$ is equal to $q \Phi$ photons/second. 

The random process describing the detections of this PF-SPAD pixel is not a
Poisson process, but can be modeled as a renewal process [\ref{sr1}] with a
shifted exponential inter-arrival distribution which has a mean $\tau_d +
\frac{1}{q\Phi}$ and variance $\frac{1}{q^2\Phi^2}$. Using the central limit
theorem for renewal processes, $N_T$ is approximately Gaussian distributed with
mean:
$$
\mathbf{E}[N_T] = \frac{q\Phi T}{1+q\Phi \tau_d}
$$ 
and variance:
$$
\mathrm{Var}[N_T] = \frac{q \Phi T}{(1+q\Phi \tau_d)^3}.
$$

\noindent\textbf{Quantization Noise:}
An additional source of variance arises due to quantization noise which we can
treat as uniformly distributed between $0$ and $1$ with variance $1/12$. The
c.d.f. of the estimated photon flux $\hat\Phi$ can be computed using the delta
method [\ref{sr2}]:

\begin{align}
  F_{\hat\Phi}(x) &= \mathsf{Pr}( \hat\Phi \leq x ) \\
  &= \mathsf{Pr}\left( \frac{1}{q}\frac{N_T}{T-\tau_d N_T}  \leq x \right) \\
  & \approx  \mathsf{Pr}\left( N_T \leq \frac{qxT}{1+qx\tau_d} \right)
                \label{eq:nonneg_approx} \\
  & = \frac{1}{2} \left( 1+\mathrm{erf}\left( \frac{\frac{qxT}{1+qx\tau_d} - \frac{q\Phi T}{1+q\Phi \tau_d}}{\sqrt{2} \sqrt{\frac{q\Phi T}{(1+q \Phi \tau_d)^3}+\frac{1}{12} }} \right) \right)
                \label{eq:gaussian_cdf}\\
  & \approx \frac{1}{2} \left( 1+\mathrm{erf}\left(\frac{x - \Phi}{\sqrt{2}\sqrt{\frac{\Phi(1+q\Phi\tau_d)}{qT}+\frac{(1+q\Phi\tau_d)^4}{12q^2 T^2}}}\right) \right) \label{eq:taylor_approx}
\end{align}
where Equation~(\ref{eq:nonneg_approx}) follows from the fact that in practice the
denominator is always non-negative since $N_T \leq \lfloor T/\tau_d \rfloor$,
Equation~(\ref{eq:gaussian_cdf}) follows from the formula for the Gaussian c.d.f. of
$N_T$ with $\mathrm{erf}$ denoting the error function [\ref{sr3}],
and Equation~(\ref{eq:taylor_approx}) is follows from a first order Taylor series
approximation. This result shows that $\hat\Phi$ is approximately normally
distributed with mean equal to the true photon flux and variance given by the
denominator in Equation~(\ref{eq:taylor_approx}).
\newline\newline
\noindent\textbf{Dark Count and Afterpulsing Bias:}
In addition to quantization and shot noise that introduce variance in the
estimated photon flux, PF-SPADs also suffer from dark counts and afterpulsing
noise that introduce a bias in the estimated flux. The dark count rate
$\Phi_\mathrm{dark}$ is often given in published datasheets and can be used as
the bias term. Afterpulsing noise is quoted in datasheets as afterpulsing
probability which denotes the probability of observing a spurious afterpulse
after the dead time $\tau_d$ has elapsed. Due to an exponentially distributed
waiting time, the probability of observing a gap between true photon-induced
avalanches is equal to $e^{-q\Phi\tau_d}$. A fraction
$p_\mathrm{ap}e^{-q\Phi\tau_d}$ of these gaps will contain afterpluses, on
average. The bias $\Delta \hat\Phi$ in the estimated flux is given by:

\begin{equation}
  \Delta \hat \Phi = \hat\Phi \frac{T}{T-N_T\tau_d} \frac{\Delta N_T}{N_T}
  =\frac{T}{T-N_T\tau_d} p_\mathrm{ap}e^{-q\Phi\tau_d}
  =q\Phi(1+\Phi\tau_d) p_\mathrm{ap}e^{-q\Phi\tau_d}. \label{eq:afterpulsing_bias}
\end{equation}

Using the bias-variance decomposition of mean-squared error, we have

\begin{equation}
  \mathsf{RMSE}(\hat\Phi) = \sqrt{ (\Phi_\mathrm{dark} + q\Phi(1+\Phi\tau_d) p_\mathrm{ap}e^{-q
  \Phi\tau_d})^2 +\frac{\Phi(1+q\Phi\tau_d)}{qT}+\frac{(1+q\Phi \tau_d)^4}{12q^2 T^2}}
\label{eq:spad_mse_approx}
\end{equation}
and the approximate closed from SNR is obtained by plugging Equation~(\ref{eq:spad_mse_approx}) into 
Equation~(\nolink{\ref{eq:snr_general_formula}}) in the main text.
\newline\newline

\clearpage
\section{Exact Formula for Numerical Computation of SNR of a SPAD
  Pixel\label{sec:suppl_note_spad_exact_snr}}
It is possible to model the exact discrete distribution of the number of counts
$N_T$ for a PF-SPAD pixel using non-asymptotic renewal theory. The times between
consecutive counts for a PF-SPAD pixel can be modeled as a shifted exponential
distribution as before. Let $X_{n}$ be the time between when the SPAD detects the
$(n-1)^{st}$ and $n^{th}$ photons ($n\geq1$). For mathematical convenience, we
assume $X_{0}=0.$ Let $F_{S_{n}}$ be the c.d.f. of the sum
$S_{n}:=\sum_{i=1}^{n}X_{n}.$ Then, by definition,
$F_{S_{n}}(T)=\mathsf{Pr}(N_T\geq n).$ Therefore we can write 

\[
p_n := \mathsf{Pr}(N_T=n)=F_{S_{n}}(T)-F_{S_{n+1}}(T)
\]
where 

\[
F_{S_{n}}(T)=1-\sum_{k=0}^{n-1}\frac{(T-n\tau_{d})^k (q\Phi)^k}{k!}\,e^{-(T-n\tau_{d})q\Phi}=1-Q(n-1,(T-n\tau_{d})q\Phi)
\]
and $Q(\cdot,\mu)$ is the c.d.f. of a Poisson random variable with rate $\mu,$
also known as the regularized gamma function [\ref{sr3}]. For
convenience, define:

\[
Q_{q,\Phi,T,\tau_{d}}(k):=Q(k,(T-k\tau_{d})q\Phi).
\]
The following formula can now be used to numerically compute the probability mass
function of $N_T$:

\begin{equation}
p_n = \begin{cases}
Q_{q,\Phi,T,\tau_{d}}(n)-Q_{q,\Phi,T,\tau_{d}}(n-1) & \mbox{for } 1\leq n\leq\lfloor\frac{T}{\tau_{d}}\rfloor\\
0 & \mbox{otherwise.}
\end{cases} \label{eq:exact_pmf}
\end{equation}
Using the bias-variance decomposition, the RMSE can be written as:

\begin{equation}
  \mathsf{RMSE}(\hat\Phi) = \sqrt{ (\Phi_\mathrm{dark} + q\Phi(1+\Phi\tau_d) p_\mathrm{ap}e^{-q\Phi\tau_d})^2 + \sum_{n=1}^{\lfloor\frac{T}{\tau_d}\rfloor}p_n \; \left(\frac{1}{q}\frac{n}{T-n\tau_d} -\Phi \right)^2}
\label{eq:spad_mse_exact}
\end{equation}
and SNR can be computed by plugging Equation~(\ref{eq:exact_pmf}) and
Equation~(\ref{eq:spad_mse_exact}) in Equation~(\nolink{\ref{eq:snr_general_formula}})
in the main text. Note that although this formula is exact, it does not lend itself
to an intuitive interpretation as the approximate formula in
Equation~(\ref{eq:spad_mse_approx}) which decomposes the sources of variance
into shot noise and quantization noise.

\clearpage
\section{Various Sources of Noise Affecting the PF-SPAD Flux Estimate
  \label{sec:suppl_note_spad_sources_of_noise}}

Various unique properties of the shot noise and quantization noise were
discussed in the main text. Another surprising result is that the effect of
afterpulsing bias first increases and then decreases with incident photon flux.
Recall that afterpulses are correlated with past avalanche events. At very low
incident photon flux there are very few photon-induced avalanches which implies
that there are even fewer afterpulsing avalanches. At very high photon flux
values, the afterpulsing noise is overwhelmed by the large number of true
photon-induced avalanches that leave negligible temporal gaps between
consecutive dead time windows. However, for most modern SPAD pixels,
afterpulsing noise is so small that it can often be ignored. The plot in
Supplementary Figure \ref{fig:spad_sources_of_noise_flux_estimate} shows an
afterpulsing error curve using an unrealistically high afterpulsing rate to
accentuate the trend as a function of incident flux.

\begin{figure}[!ht]
  \centering \includegraphics[width=0.5\textwidth]{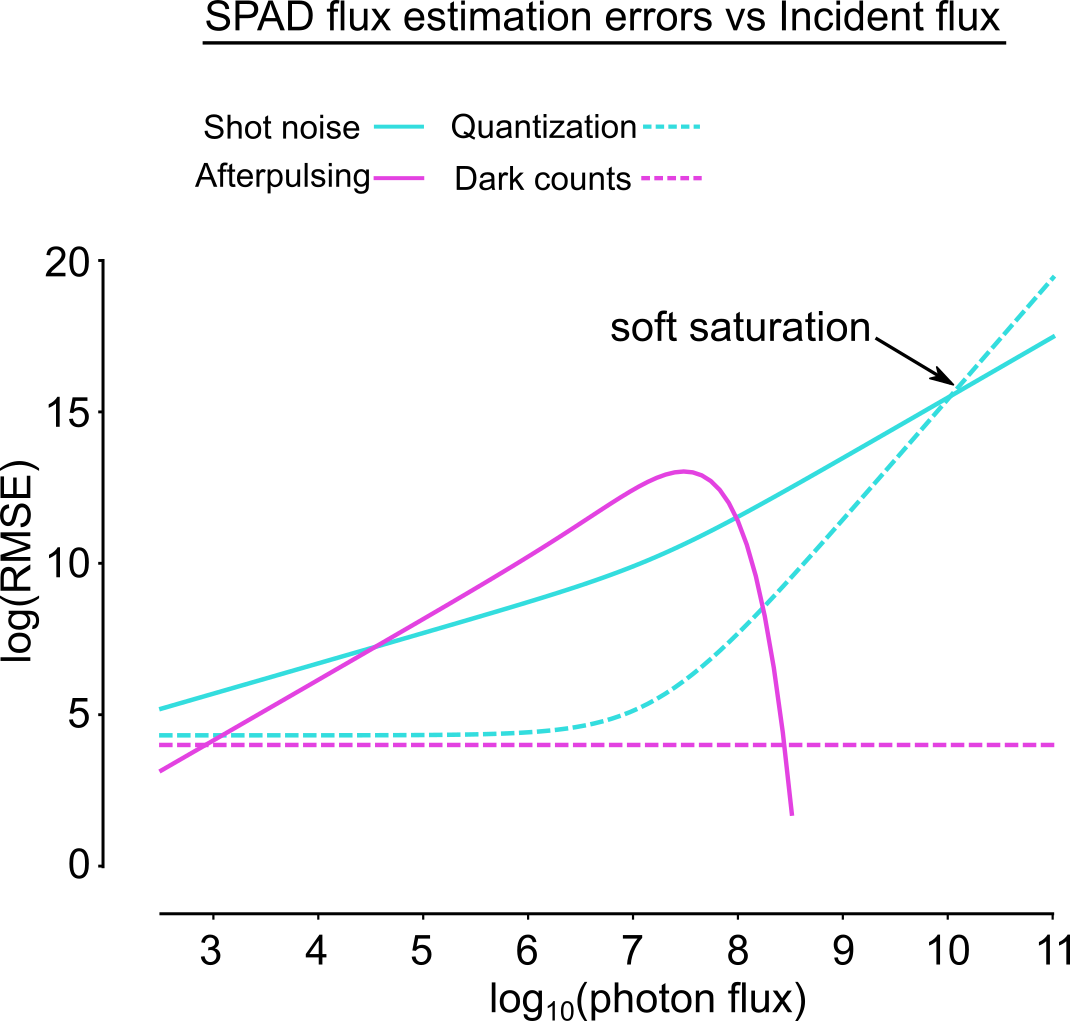}
  \caption{\textbf{Effect of various sources of noise on the estimated photon flux
  for a conventional and a SPAD pixel.} This figure shows the contributions to
  the flux estimation error from various sources of noise in a SPAD pixel.
  Quantization noise and shot noise were discussed in the main text and in
  Figure~\ref{fig:snr}. Bias due to afterpulsing noise increases with incident
  flux and then decreases. Dark count noise remains small and constant at all
  flux levels. In order to accentuate the trend of afterpulsing error with
  incident flux, this plot uses an unrealistically high afterpulsing
  probability of 30\%, which is much higher than the 1\% probability for our
  hardware prototype. \label{fig:spad_sources_of_noise_flux_estimate}}
\end{figure}

\clearpage
\section{SNR of a Conventional Sensor Pixel\label{sec:suppl_note_ccd}}
A conventional CCD or CMOS pixel suffers from a hard saturation limit due to its full
well capacity, $N_\mathrm{FWC}$. Assuming a quantum efficiency of $0<q<1$, an incident
photon flux of $\Phi$ photons/second and an exposure time $T$ seconds, the photon
counts $N_T$ follow a Gaussian distribution with mean $q\Phi T$ and variance 
$q \Phi T + \sigma_r^2$ where $\sigma_r$ is the read noise of the pixel.
The estimated flux is given by [\ref{sr8}]:

$$
\hat \Phi_\text{CCD} = \begin{cases}
  \frac{N_T}{qT}, & N_T < N_\mathrm{FWC} \\
  \infty,         & N_T = N_\mathrm{FWC}.
\end{cases}
$$
The RMSE of the estimated flux is given by:

$$
\mathsf{RMSE}(\hat \Phi_\text{CCD}) = 
\sqrt{\mathbf{E}[(\hat\Phi_\text{CCD}-\Phi)^{2}]} =
\begin{cases}
  \frac{\sqrt{q \Phi T + \sigma_r^2}}{q T}, & \Phi < \frac{N_\mathrm{FWC}}{qT} \\
  \infty, & \Phi \geq \frac{N_\mathrm{FWC}}{qT}.
\end{cases}
$$
which leads to the following formula for SNR of conventional pixel:

\begin{equation}
\mathsf{SNR}_\textrm{CCD}(\Phi) = \begin{cases}
  10 \log_{10} \left( \frac{q^2 \Phi^2 T^2}{q \Phi T + \sigma_r^2} \right), & \Phi < \frac{N_\mathrm{FWC}}{qT} \\
  -\infty, & \Phi \geq \frac{N_\mathrm{FWC}}{qT}.
\end{cases} \label{eq:conventional_sensor_snr}
\end{equation}
This formula does not account for dark current noise because it is only relevent
at extremely low incident photon flux values with very long exposure times of
many minutes or longer.

\clearpage
\section{Effect of Varying Exposure Time \label{sec:suppl_note_varying_exp_time}}
The notions of quantum efficiency and exposure time are interchangeable in case
of conventional image sensors; Equation~(\ref{eq:conventional_sensor_snr}) remains
unchanged if the symbols $q$ and $T$ were to be swapped. This is not true for a
PF-SPAD sensor where changing $q$ and changing $T$ has different effects on the
overall SNR. This is because the SPAD pixel has an asymptotic saturation limit
of $T/\tau_d$ counts which is a function of exposure time, unlike a
conventional sensor whose full well capacity is a fixed constant independent of
exposure time. As shown in Supplementary
Figure~\ref{fig:snr_effect_of_parameters}, decreasing exposure time decreases the
maximum achievable SNR value of a SPAD sensor. Experimental results were
obtained from our hardware prototype using a dead time of
$\SI{300}{\nano\second}$ and capturing photon counts with two different
exposure times of $\SI{0.5}{\milli\second}$ and $\SI{5}{\milli\second}.$

\begin{figure}[!ht]
\centering \includegraphics[width=0.9\textwidth]{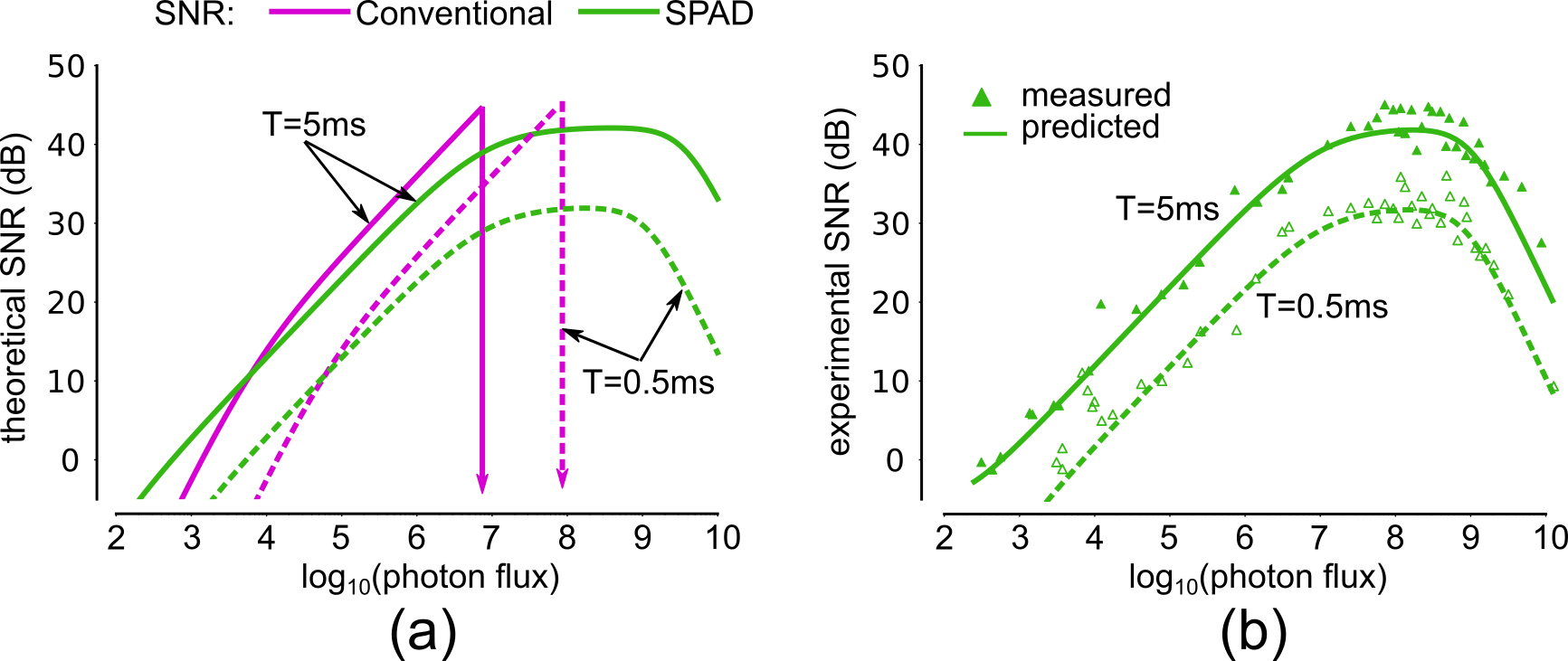}
  \caption{\textbf{Effect of varying exposure time on SNR}
  (a) For a conventional sensor, decreasing exposure time translates the SNR
  curve towards higher photon flux values while keeping the overall shape of
  the curve same. However, for a PF-SPAD pixel, decreasing exposure time decreases
  the maximum achievable SNR.  (b) Experimental SNR data obtained with two
  exposure times. The SNR curves decay more rapidly than (a) due to additional
  dead time uncertainty effects in our hardware prototype, but the decrease in
  maximum achievable SNR is still clearly seen.
  \label{fig:snr_effect_of_parameters}}
\end{figure}

\clearpage
\section{Details of SPAD Simulation Model and Experimental Setup\label{sec:suppl_simulation_model}}
We implemented a time-domain simulation model for a PF-SPAD pixel to validate our
theoretical formulas for the PF-SPAD response curve and SNR.  Photons impinge the
simulated PF-SPAD pixel according to a Poisson process; a fraction of these
photons are missed due to limited quantum efficiency. The PF-SPAD pixel counts an
incident photon when it arrives outside a dead time window. The simulation
model also accounts for spurious detection events to dark counts and
afterpulsing. The pseudo-code is shown in Supplementary
Figure~\ref{alg:spad_simulation_model}. 

\paragraph*{PF-SPAD and Conventional Sensor Specifications}
Each pixel in the simulated PF-SPAD array mimics the specifications of the
single-pixel hardware prototype. Each pixel in our simulated conventional
sensor array uses slightly higher specifications than the one we used in our
experiments. It has a full well capacity of 33,400 electrons, quantum efficiency
of 90\% and read noise of 5 electrons. 

\begin{figure}[!ht]
  \centering \includegraphics[width=1.0\textwidth]{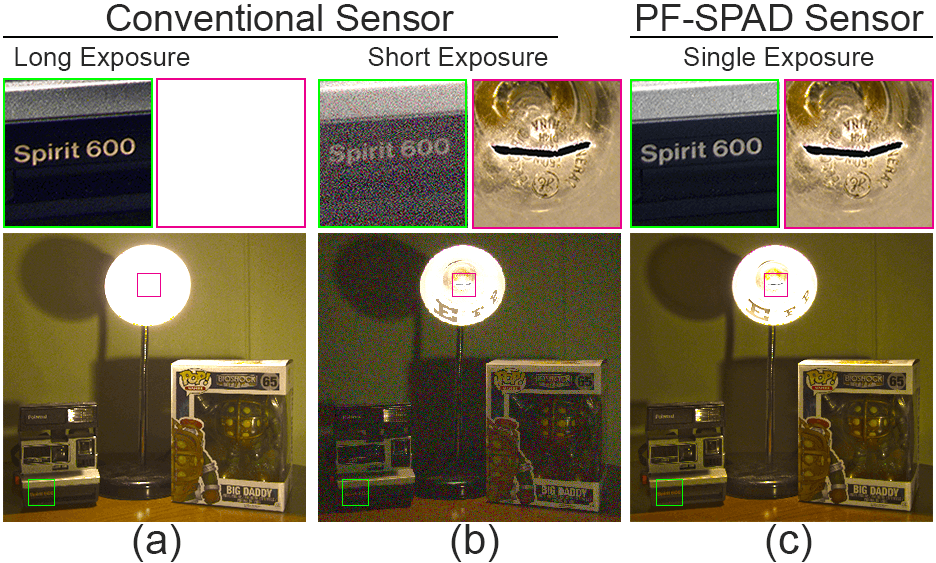}
  \caption{\textbf{Simulated indoor HDR scene.} (a) Long exposure capture using a
  conventional camera captures darker regions of the scene but the regions
  around the bulb are completely saturated. (b) Short exposure time capture
  using a conventional sensor shows details of bulb filament but darker regions
  of the scene appear grainy due to underexposure. (c) A single capture using a
  PF-SPAD captures both bright and dark regions simultaneously. The original HDR
  image was captured using a Canon EOS Rebel T5 DSLR camera with 10 stops and
  rescaled to cover $10^6:1$ dynamic range.
  \label{fig:simulated_indoor_lcpr}}
\end{figure}

The single-pixel SPAD simulator was used for generating synthetic color images
from a hypothetical megapixel SPAD array camera. The ground truth photon flux
values were obtained from an exposure bracketed HDR image that covered over 10
orders of magnitude in dynamic range. Unlike regular digital images that use 8
bit integers for each pixel value, an HDR image is represented using floating
point values that represent the true scene radiance at each pixel. These
floating point values were appropriately scaled and used as the ground-truth
photon flux to generate a sequence of photon arrival times following Poisson
process statistics. Red, green and blue color channels were simulated
independently. Results of simulated HDR images are shown in Supplementary
Figures~\ref{fig:simulated_observatory}, \ref{fig:simulated_sunset_scene} and
\ref{fig:simulated_indoor_lcpr}.

\begin{figure}[!ht]
\hrulefill
  \begin{algorithmic}[1]
    \Require $\;\,\Phi$: true incident photon flux \par
             $T$: exposure time \par
             $q$: SPAD pixel photon detection probability (quantum efficiency) \par
             $\Phi_\mathrm{dark}$: SPAD dark count rate \par
             $\tau_d$: dead time \par
             $p_\mathrm{ap}$: afterpulsing probability \par
    \Ensure  $N_T$: number of photon detections
  
    \Procedure {PFSPADSimulator}{$\Phi, T, q, \tau_d, p_\mathrm{ap}$}
    \State Reset number of photon detections $N_T \gets 0$
    \State Reset last detection time $t_\mathrm{last} \gets -\infty$
    \State Reset simulation time $t \gets 0$
    \State Initialize afterpulse time-stamp array $\mathbf{t}_\mathrm{ap}=[\,]$
    \While {$t \leq T$}
      \State Process timestamps in the after-pulsing time vector $\mathbf{t}_\mathrm{ap}$
      \State Generate next photon time-stamp $t \gets t+\mathrm{Exp}(q\Phi+\Phi_\mathrm{dark})$
      \If {$t \geq t_\mathrm{last}+\tau_d$}
        \State Append next afterpulse time-stamp to $\mathbf{t}_\mathrm{ap}$ 
        \State $N_T \gets N_T+1$
        \State $t_\mathrm{last} \gets t$
      \EndIf
    \EndWhile
    \EndProcedure
  \end{algorithmic}
  \hrulefill
  \caption{\textbf{Computational model of a PF-SPAD pixel.} \label{alg:spad_simulation_model}}
\end{figure}

\begin{figure}[!ht]
  \centering \includegraphics[width=1.0\textwidth]{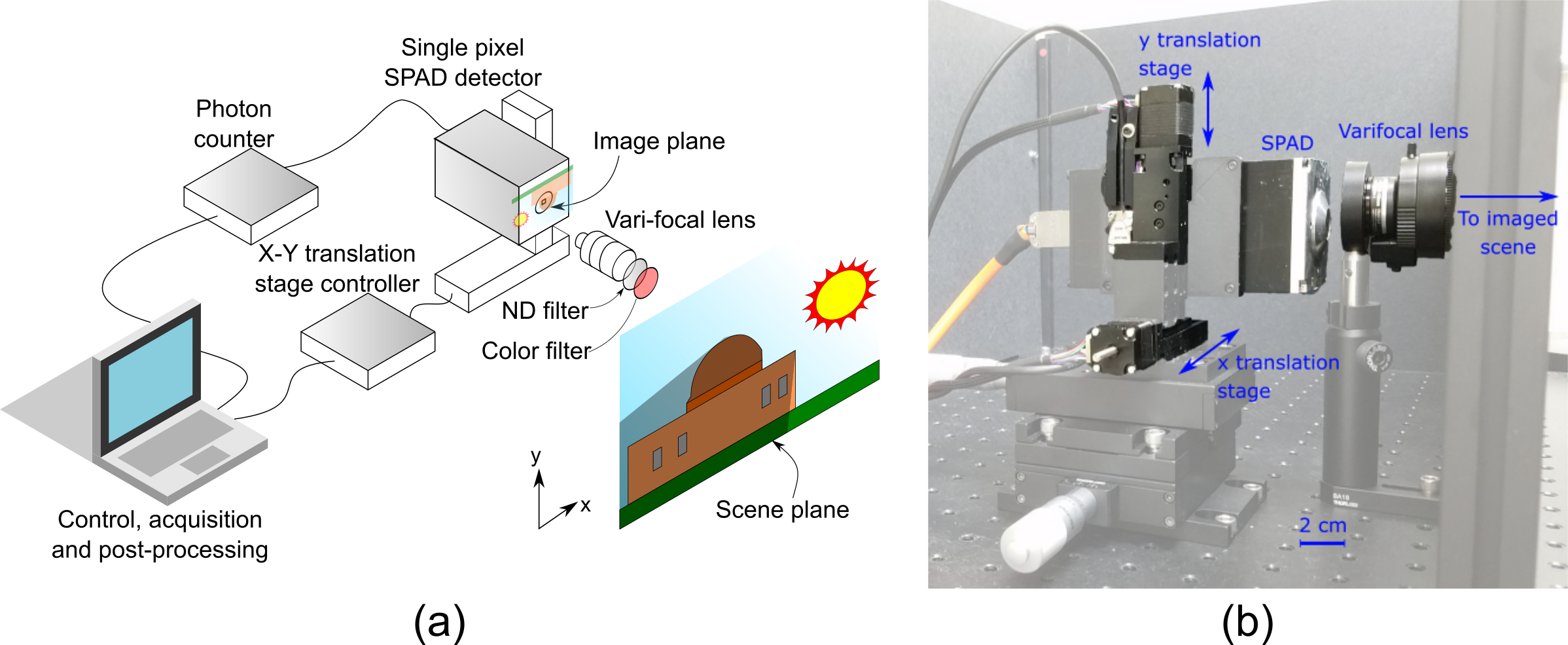}
  \caption{\textbf{Experimental setup for raster scanning with
      a single-pixel PF-SPAD sensor.} (a) The setup consists of a SPAD module
      mounted on two translation stages, and a variable focal length lens that
      relays the imaged scene onto the image plane. Photon counts are captured
      using a free-running time-correlated single-photon counting module
      operated without a synchronization signal. (b) A picture of our SPAD
      sensor mounted on the translatation stages.
  \label{fig:expt_setup}}
\end{figure}

\section*{Experimental Setup Details}
The single-pixel SPAD from our hardware prototype has a pitch of
$\SI{25}{\micro\meter}$, quantum efficiency of 40\%, dark count rate of 100
photons/second and 1\% afterpulsing probability. The dead time is programmable
and was set to $\SI{149.7}{\nano\second}$ and exposure time to
$\SI{5}{\milli\second}$. This corresponds to an asymptotic saturation limit of
33,400 photons. 

Each pixel in our machine vision camera (Point Grey GS3-U3-23S6M-C) has a full
well capacity of 32,513 electrons, a peak quantum efficiency of 80\% and a
Gaussian-distributed read noise with a standard deviation of 6.83 electrons.
Note that the asymptotic saturation limit of the PF-SPAD pixel is similar to
the full well capacity of this machine vision camera to enable fair comparison.

\clearpage 
\section{Effect of Dead Time Jitter\label{sec:suppl_note_spad_dead_time_unc}}
In practice the dead time window is controlled using digital timer circuits
that have a limited precision dictated by the clock speed. The hardware used in
our experiments has a clock speed of 167 MHz which introduces a variance of
\SI{6}{\nano\second} in the duration of the dead time window. As a result the
dead time $\tau_d$ can no longer be treated as a constant but must be treated
as a random variable $T_d$ with mean $\mu_d$ and variance $\sigma_d.$ The
inter-arrival distribution in Equation~(\ref{eq:interarrival_distr}) must be understood
as a conditional distribution, conditioned on $T_d=\tau_d$. The mean and
variance of the time between photon detections can be computed using the law of
iterated expectation [\ref{sr1}]:

$$\mathbf{E}[X_n] = \mathbf{E}[\mathbf{E}[X_n|T_d]] = \mu_d+\frac{1}{q\Phi}$$ 
and 

$$\mathrm{Var}[X_n] =
\mathbf{E}[(X_{n}-\mathbf{E}[X_{n}])^{2}]=\mathbf{E}[\mathbf{E}[(X_{n}-\mathbf{E}[\mathbf{E}[X_{n}|\mathrm{T}_{d}]])^{2}|\mathrm{T}_{d}]]
= 1/q^2 \Phi^2 + \sigma_d^2.$$

\begin{figure}[!ht]
\centering \includegraphics[width=0.5\textwidth]{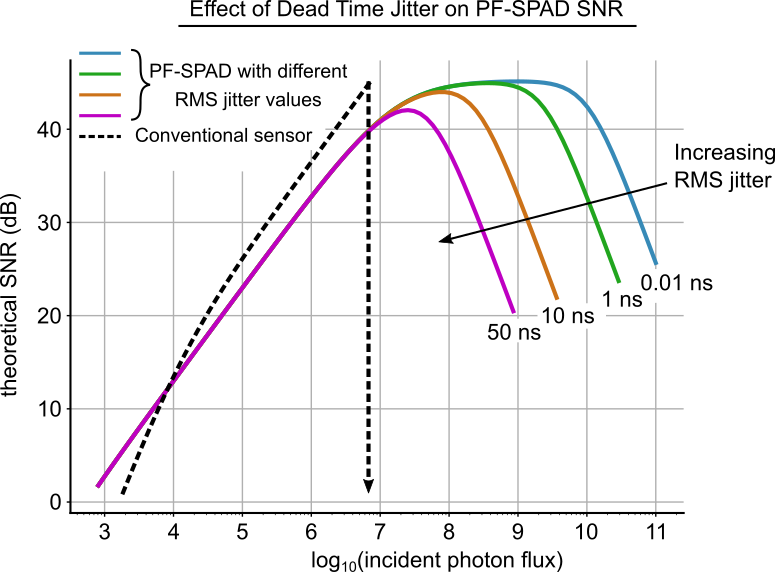}
\caption{{\bf Effect of dead time jitter on PF-SPAD SNR} This figure shows
theoretical effect of different values of dead time jitter on the PF-SPAD's SNR
is shown. (5 ms exposure time, 149.7 ns dead time, 40\% quantum efficiency and
100 Hz dark count rate.) \label{fig:dead_time_jitter}}
\end{figure}

Using similar computations as those leading to Equation~(\ref{eq:taylor_approx}), we can
derive a modified shot noise variance term equal to $\frac{\Phi (1+ q^2\Phi^2
\sigma_d^2) (1+q\Phi \mu_d)}{qT}$ that must be used in Equation~(\ref{eq:spad_mse_approx})
to account for dead time variance. All instances of $\tau_d$ in
Equation~(\ref{eq:spad_mse_approx}) must be replaced by its mean value $\mu_d$. Supplementary
Figure~\ref{fig:dead_time_jitter} shows theoretical SNR curves for a PF-SPAD pixel with a nominal
dead time duration of \SI{149.7}{\nano\second}. Observe that the 30~dB dynamic
range degrades by almost 3 orders of magnitude when the dead time jitter
increases from \SI{0.01}{\nano\second} to \SI{50}{\nano\second}. For reference, our hardware
prototype has a dead time jitter of \SI{6}{\nano\second} RMS.

\clearpage
\section{Comparison with Quanta Image Sensors\label{sec:suppl_note_qis_snr}}
A quanta image sensor (QIS) [\ref{sr4},\ref{sr5},\ref{sr6}] improves dynamic
range by spatially oversampling the 2D scene intensities using sub-diffraction
limit sized pixels called \emph{jots}. Each jot has a limited full well
capacity, usually just one photo-electron. The PF-SPAD imaging modality
presented in this paper is different from these methods. Instead of using a
SPAD as a binary pixel [\ref{sr4}] and relying on spatial oversampling, the
PF-SPAD achieves dynamic range compression by allowing the dead time windows to
shift randomly based on the most recent photon detection time and performing
adaptive photon rejection. This is equivalent to the ``event-driven recharge''
method described in [\ref{sr7}].

We now derive the image formation model and an expression for SNR for a QIS and
other related methods that use equi-spaced time bins [\ref{sr6}], and show
that their dynamic range is lower than what can be achieved using a PF-SPAD.

\subsection*{QIS Image Formation and Flux Estimator}
The output response of a QIS is logarithmic and mimics silver halide
photographic film [\ref{sr5}]. Each jot has a binary output, and the final
image is formed by spatio-temporally combining groups of jots called a
``jot-cube''. Let $\tau_b$ be the temporal bin width and for mathematical
convenience, assume that the exposure time $T$ is an integer multiple of
$\tau_b$, so that there are $N = T/\tau_b$ uniformly spaced time bins that
split the total exposure duration. Suppose the jot-cube is exposed to a
constant photon flux of $\Phi$ photons/second, and each jot has a quantum
efficiency $0<q<1$. The number of photons received by each jot in a time
interval $\tau_b$ follows a Poisson distribution with mean $q\Phi\tau_b$.
Therefore the probability that the binary output of a jot is $0$ is given by:

$$
\mathsf{Pr}(\mathrm{jot=0}) = e^{-q\Phi\tau_b}
$$
and the probability that the binary output of a jot is $1$ is given by the probability 
of observing $1$ or more photons:

$$
\mathsf{Pr}(\mathrm{jot=1}) = 1-e^{-q\Phi\tau_b}.
$$

Let $N_T$ denote the total photon counts output from a jot-cube with $N$ jots.
Then $N_T$ follows a binomial distribution given by:

$$
\mathsf{Pr}(N_T=k) = \binom{N}{k}(1-e^{-q\Phi\tau_b})^k (e^{-q\Phi\tau_b})^{N-k},
$$
for $0\leq k\leq N$. The maximum-likelihood estimate of the photon flux is given by:

\begin{equation}
  \hat\Phi_\textrm{QIS} = \frac{1}{q\tau_b}\log\left(\frac{T}{T-N_T\tau_b}\right).
\label{eq:qis_flux_from_counts}
\end{equation}

Our PF-SPAD flux extimator has a higher dynamic range than this uniform binning
method. This can be intuitively understood by noting that in the limiting case
of $\tau_b = \tau_d$ both schemes have an upper limit on photon counts given by
$N_T \leq T/\tau_d,$ but the QIS estimator in
Equation~(\ref{eq:qis_flux_from_counts}) saturates and flattens out more
rapidly than the PF-SPAD estimator in
Equation~(\nolink{\ref{eq:spad_flux_from_counts}}) from the main text:
$$
\frac{d\hat\Phi_\textrm{QIS}}{d N_T} = \frac{1}{q} \frac{1}{T - N_T\tau_b}
   <  \frac{1}{q} \frac{T}{(T-N_T\tau_d)^2} \label{eq:spad_qis_inequality} 
   = \frac{d\hat\Phi}{d N_T}.
$$

\subsection*{SNR of a QIS}
The variance of the QIS flux estimator can be computed numerically using the
binomial probability mass function of $N_T$. For convenience, a closed form
expression can be derived using a Gaussian approximation, similar to the
approximation techniques used for deriving the SNR of a PF-SPAD pixel in
Equation~(\ref{eq:taylor_approx}). The Gaussian approximation to a binomial
distribution suggests $N_T$ has a normal distribution with mean
$N\,(1-e^{-q\Phi\tau_b})$ and variance
$N\,e^{-q\Phi\tau_b}(1-e^{-q\Phi\tau_b})$. Next, the c.d.f. of the estimated
flux is given by:

\begin{align}
  F_{\hat\Phi_\textrm{QIS}}(x) &= \mathsf{Pr}(\hat\Phi_\textrm{QIS} \leq x) \nonumber \\
  &= \mathsf{Pr}\left( -\frac{1}{q\tau_b}\log\left(1 - \frac{N_T}{N}\right) \leq x\right) \nonumber \\
  &= \mathsf{Pr}(N_T \leq (1-e^{-xq\tau_b})N ) \nonumber \\
  &= \frac{1}{2}\left(1+\mathrm{erf}\left(\frac{N(1-e^{-xq\tau_b})-N(1-e^{-\Phi q\tau_b})}{\sqrt{2}\sqrt{N(1-e^{-\Phi q\tau_b})e^{-\Phi q\tau_b}}}\right)\right) \label{eq:qis_normal_1} \\
  &\approx \frac{1}{2}\left(1+\mathrm{erf}\left(\sqrt{N}\frac{(x-\Phi)e^{-\Phi q\tau_b}q\tau_b}{\sqrt{2(1-e^{-\Phi q\tau_b})e^{-\Phi q\tau_b}}}\right)\right) \label{eq:qis_taylor}\\
  &= \frac{1}{2}\left(\mathrm{1+erf}\left(\frac{x-\Phi}{\sqrt{2}\sqrt{\frac{(1-e^{-\Phi q\tau_b})}{q^{2}T\tau_b e^{-\Phi q\tau_b}}}}\right)\right) \label{eq:qis_normal_2}
\end{align}
where Equation~(\ref{eq:qis_normal_1}) follows from the formula for the c.d.f. of a
Gaussian distribution and Equation~(\ref{eq:qis_taylor}) is obtained after making a
Taylor series approximation.  The final form of Equation~(\ref{eq:qis_normal_2})
suggests that the estimated photon flux follows a normal distribution with
variance $\frac{(1-e^{-\Phi q\tau_b})}{q^{2}T\tau_b e^{-\Phi q\tau_b}}.$

The read noise of each jot affects the RMSE of the QIS sensor at low incident
flux.  We note that at low flux values there are, on average, $q\Phi T$
bins already filled by true photon counts leaving $N-q\Phi T$ bins empty. Read noise
will cause some of these empty bins to contain false positives and introduce
additional noisy counts equal to 
$\frac{1}{2}(N-q\Phi T)
\left(1+\mathrm{erf}\left(\frac{1}{2\sqrt{2}\sigma_r}\right)\right)$,
where $\sigma_r$ is the read noise
standard deviation. This corresponds to a bias of
$\frac{1}{2}(\frac{1}{q\tau_b}-\Phi)
\left(1-\mathrm{erf}\left(\frac{1}{2\sqrt{2}\sigma_r}\right)\right)$ 
in the estimated photon flux. Incorporating this bias term together with the 
variance associated with the Gaussian distribution of the estimated photon flux,
the RMSE is given by
$$
\mathsf{RMSE}(\hat\Phi_\textrm{QIS}) = \sqrt{\max\left\{ 0,\frac{1}{2} \left(\frac{1}{q\tau_b}-
\Phi\right)\left(1-\mathrm{erf}\left(\frac{1}{2\sqrt{2}\sigma_{r}}\right)\right)
\right\}^{2}+\frac{1-e^{-q\Phi\tau_b}}{q^{2}T\tau_b e^{-q\Phi\tau_b}}}.
$$

\begin{figure}[!ht]
  \centering \includegraphics[width=0.5\textwidth]{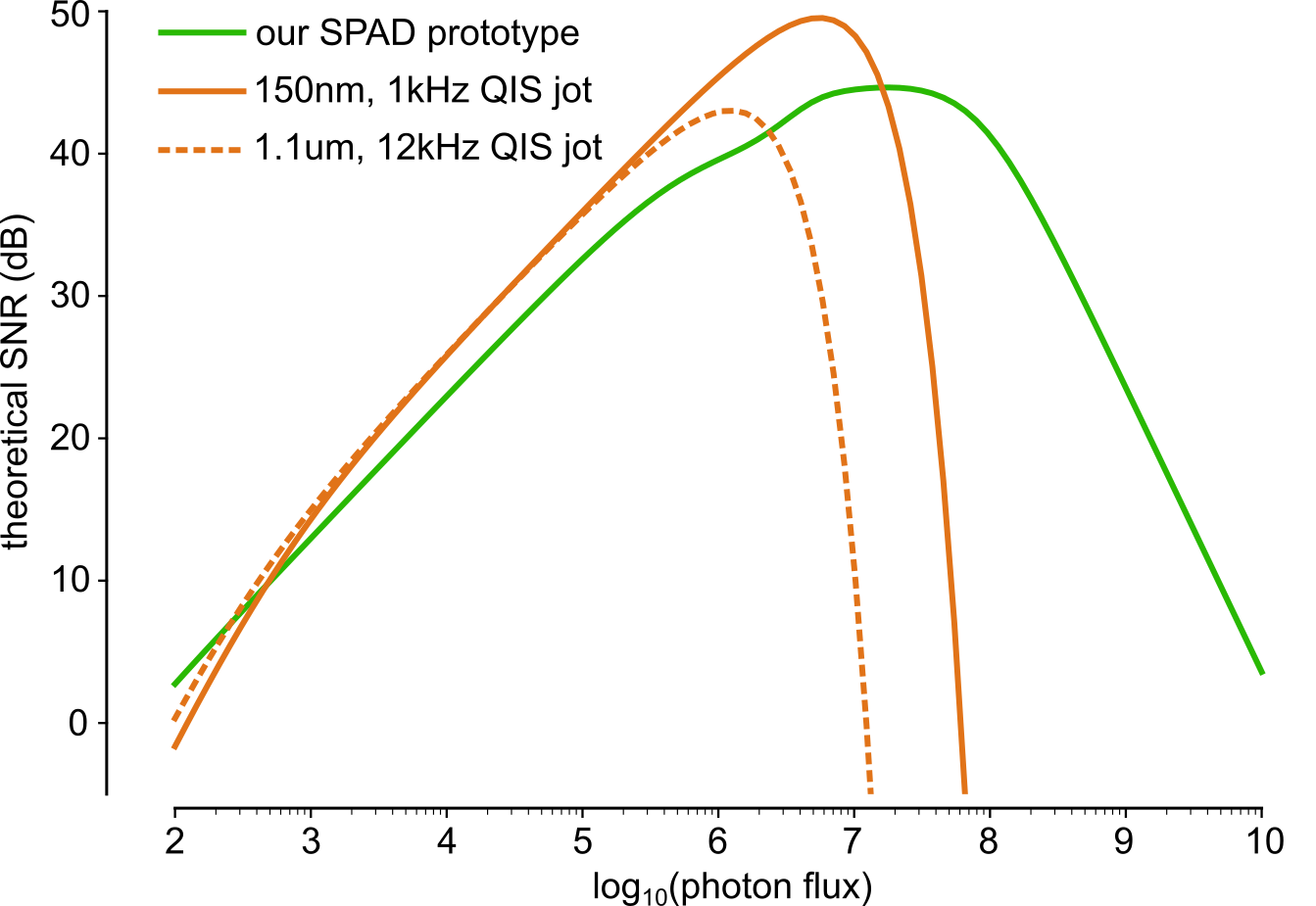}
  \caption{\textbf{Theoretical SNR curves for a SPAD pixel compared to the
  effective SNR of a QIS jot block occupying the same area as the SPAD pixel.}
  Each QIS jot has a read noise standard deviation of 0.13 electrons and
  quantum efficiency of 80\%. The SPAD pixel has a dead time of
  \SI{150}{\nano\second}, dark count rate of 100 photons/s, 40\% quantum
  efficiency and 1\% afterpulsing rate. A fixed exposure time of
  \SI{5}{\milli\second} is assumed for both types of pixels.  Sub-diffraction
  limit jot sizes of under \SI{150}{\nano\meter} will be required to obtain
  similar dynamic range as a single \SI{25}{\micro\meter} SPAD pixel.
  \label{fig:qis_vs_spad_snr}}
\end{figure}

A single jot only generates a binary output and must be combined into a
jot-cube to generate the final image. One way to obtain a fair comparison
between a PF-SPAD pixel and a jot-cube is computing the SNR for a fixed image
pixel size and fixed exposure time. We use a square grid of jots that spatially
occupy the same area as our single PF-SPAD pixel that has a pitch of
\SI{25}{\micro\meter}. Supplementary Figure~\ref{fig:qis_vs_spad_snr} shows the
SNR curves obtained using our theoretical derivations for a QIS jot-cube and a
single PF-SPAD pixel. State of the art jot arrays are limited to a pixel size
of around $\SI{1}{\micro\meter}$ and frame rates of a few kHz. These SNR curves
show that we will require large spatio-temporal oversampling factors and
extremely small jots to obtain similar dynamic range as a single PF-SPAD pixel.
For example, a \SI{150}{\nano\meter} jot size can accommodate almost 30,000
jots in a $25\times$\SI{25}{\micro\meter}$^2$ area occupied by the PF-SPAD
pixel and can provide similar dynamic range and higher SNR than our PF-SPAD
pixel when operated at a frame rate of 1 kHz. QIS technology will require an
order of magnitude increase in frame readout rate or an order of magnitude
reduction in jot size to bring it closer to the dynamic range achievable with
our PF-SPAD prototype.

\clearpage 
\section{Additional Simulated and Experimental Results}

\begin{figure}[!ht]
  \centering \includegraphics[width=1.0\textwidth]{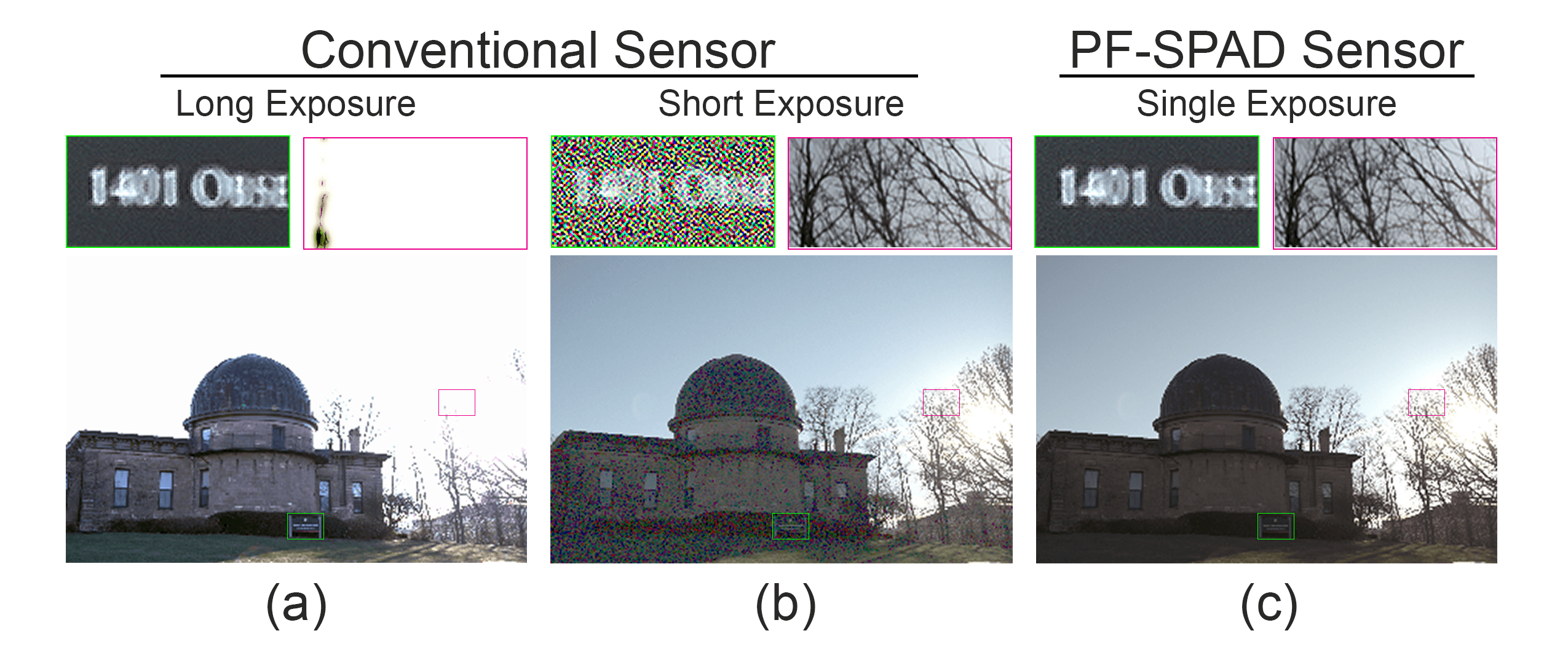}
  \caption{\textbf{Simulation-based comparison of a conventional
  image sensor and a PF-SPAD on a high dynamic range scene.} The
  ground truth high dynamic range image was obtained using a DSLR camera with
  exposure bracketing over 10 stops. (a) Simulated $\SI{5}{\milli\second}$
  exposure image of the scene obtained using a conventional camera sensor. (b)
  Simulated $\SI{50}{\micro\second}$ exposure time image using a conventional
  camera. (c) Simulated SPAD image of the same scene acquired with a single
  $\SI{5}{\milli\second}$ exposure captures the full dynamic range in a single
  shot. Identical tone-mapping was applied to all images and zoomed insets for
  a fair comparison and reliable visualization of the entire dynamic range.
  \label{fig:simulated_observatory}}
\end{figure}

\begin{figure}[!ht]
  \centering \includegraphics[width=1.0\textwidth]{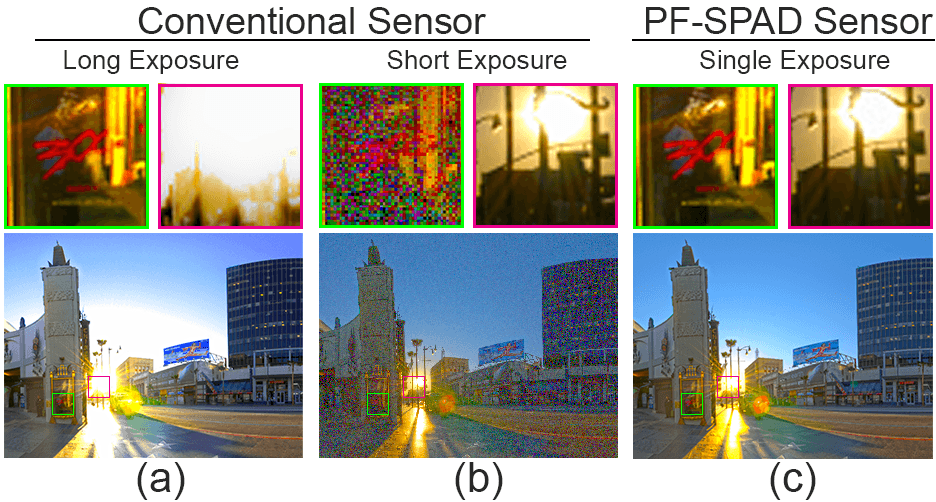}
  \caption{\textbf{Simulated outdoor HDR scene.} (a) Long exposure capture using a
  conventional camera captures darker regions of the scene but the regions
  around the sun are saturated. (b) Short exposure time capture using a
  conventional sensor prevents the sun-lit region from appearing saturated but
  results in lost information in the shadows. (c) A single capture using a
  simulated PF-SPAD array circumvents the problem of low dynamic range
  by simultaneously capturing
  both highlights and shadows. Original HDR image was obtained from
  the sIBL datasets website \texttt{www.hdrlabs.com/sibl/archive.html}.
  \label{fig:simulated_sunset_scene}}
\end{figure}

\begin{figure}[!ht]
  \centering \includegraphics[width=1.0\textwidth]{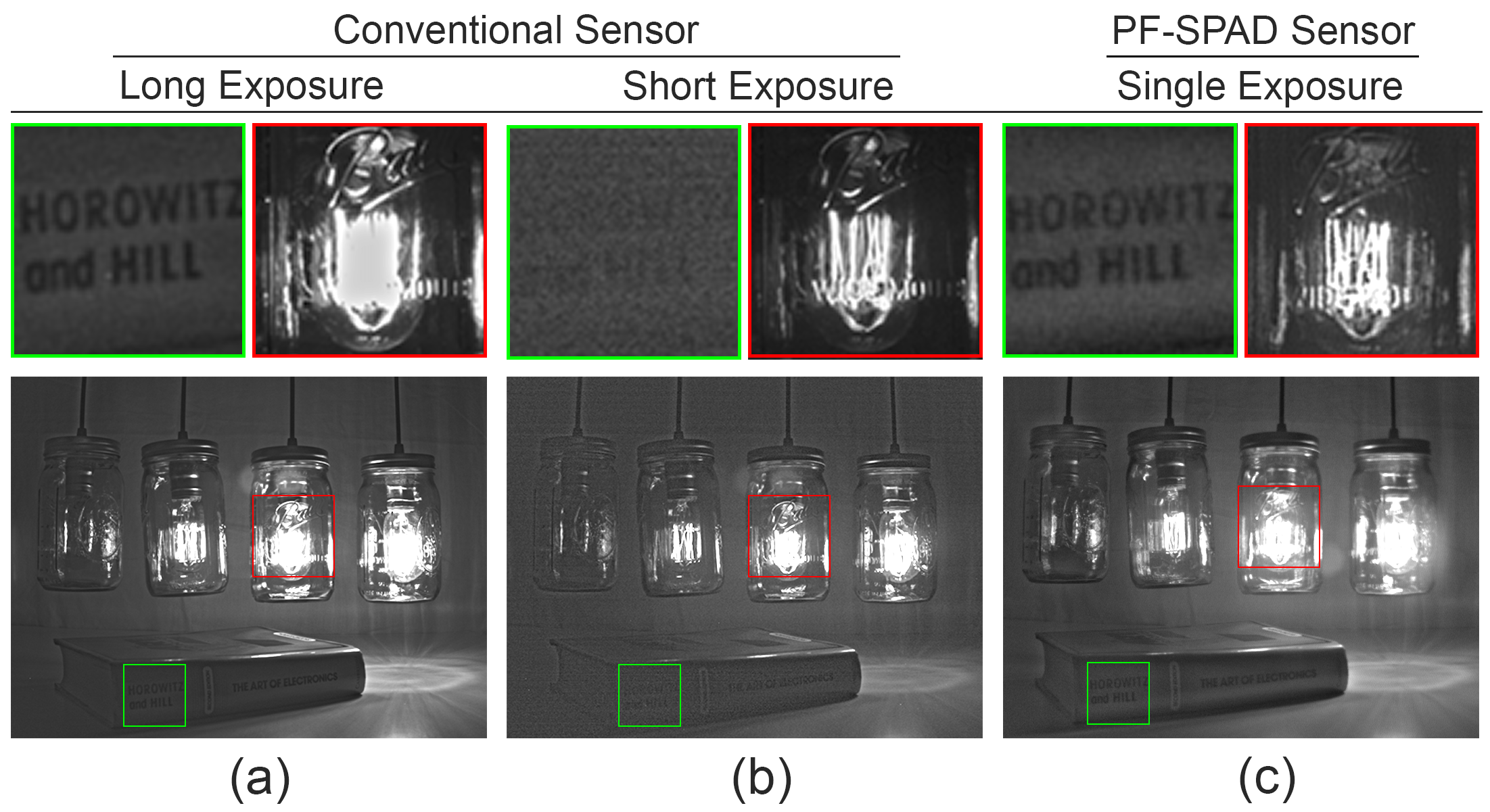}
  \caption{\textbf{Comparison of the dynamic range of images captured using a
  conventional camera and our PF-SPAD hardware prototype.}
  (a) Long exposure (\SI{5}{\milli\second}) shot using a conventional camera
  captures darker regions of the scene such as the text but the regions around
  the bulb filaments appear saturated.  (b) Short exposure time
  (\SI{0.5}{\milli\second}) capture using a shows filaments of all bulbs but
  leaves the darker part of the scene such as the book underexposed. (c) A
  single \SI{5}{\milli\second} exposure shot using the SPAD prototype captures
  the entire dynamic range. The bright bulb filament and dark text on the book
  are simultaneously visible.
  \label{fig:artistic_scene_2}}
\end{figure}

\clearpage\newpage

\section*{Supplementary References}
\renewcommand*\labelenumi{[\theenumi]}
\begin{enumerate}
    \item Abramowitz, M. \& Stegun, I. A. \emph{Handbook of Mathematical
      Functions: With Formulas, Graphs, and Mathematical Tables}, vol. 55
      (Dover American Nurses Association Publications, 1964), 9 edn. \label{sr3}
    \item Antolovic, I. M., Bruschini, C. \& Charbon, E. Dynamic range
      extension for photon counting arrays. \emph{Optics Express} \textbf{26},
      22234--22248 (2018). \label{sr7}
    \item Casella, G. \& Berger, R. L. \emph{Statistical Inference} (Pacific
      Grove, CA: Duxbury/Thomson Learning, 2002), 2nd edn. Sec. 5.5.4. \label{sr2}
    \item Dutton, N. A. W. \emph{et al.} A SPAD-based QVGA image sensor for
        single-photon counting and quanta imaging. \emph{IEEE Transactions on
        Electron Devices} \textbf{63}, 189--196 (2016). \label{sr4}
    \item Fossum, E., Ma, J., Masoodian, S., Anzagira, L. \& Zizza, R. The
      quanta image sensor: Every photon counts. \emph{Sensors} \textbf{16},
      1260 (2016). \label{sr5}
    \item Grimmett, G. R. \& Stirzaker, D. R. \emph{Probability and Random
      Processes} (Oxford University Press, 2001), 3rd edn. \label{sr1}
    \item Hasinoff, S. W. et al. Noise-Optimal Capture for High Dynamic Range
      Photography. Proc. 23rd  IEEE Conference on Computer Vision and Pattern
      Recognition (CVPR), pp. 553--560 (2010). \label{sr8}
    \item Itzler, M. A. Apparatus comprising a high dynamic range
      single-photon passive 2D imager and methods therefor (2017). \label{sr6}
\end{enumerate}

\end{document}